\documentclass[journal]{IEEEtran}

\usepackage{graphicx}
\usepackage{subcaption}

\usepackage{times,amsmath,epsfig}
\usepackage{textcomp}
\usepackage{gensymb}
\usepackage[utf8]{inputenc}
\usepackage{algorithm}
\usepackage{algpseudocode}
\newsavebox{\imgbox}
\usepackage{hyperref}
\hypersetup{
    colorlinks=true,
    linkcolor=blue,
    filecolor=magenta,      
    urlcolor=cyan,
}
\usepackage{color}

\usepackage{epstopdf}
\usepackage{cite}
\usepackage{amsmath,amssymb}

\begin{document}

\title{Decentralized Frequency Alignment\\for Collaborative Beamforming\\in Distributed Phased Arrays}

\author{Hassna Ouassal, Ming Yan, and Jeffrey A. Nanzer,~\IEEEmembership{Senior Member,~IEEE}%

\thanks{This material is based upon work supported  The Defense Advanced Research Projects Agency (grant number N66001-17-1-4045) and by the National Science Foundation (grant number DMS-1621798). The views, opinions, and/or findings contained in this article are those of the author and should not be interpreted as representing the official views or policies, either expressed or implied, of the Defense Advanced Research Projects Agency or the Department of Defense.}
\thanks{H. Ouassal and J. Nanzer are with the Department of Electrical and Computer Engineering, Michigan State University, East Lansing, MI 48824 USA (email: nanzer@msu.edu).}
\thanks{M. Yan is with the Department of Computational Mathematics, Science and Engineering and the Department of Mathematics, Michigan State University, East Lansing, MI 48824 (email:myan@msu.edu).}
}

\markboth{}%
{Shell \MakeLowercase{\textit{et al.}}: Bare Demo of IEEEtran.cls for IEEE Journals}

\maketitle

\begin{abstract}
A new approach to distributed syntonization (frequency alignment)  
for the coordination of nodes in open loop coherent distributed antenna arrays to enable distributed beamforming is presented. This approach makes use of the concept of consensus optimization among nodes without requiring a centralized control. Decentralized frequency consensus can be achieved through iterative frequency exchange among nodes. We derive a model of the signal received from a coherent distributed array and analyze the effects on beamforming of phase errors induced by oscillator frequency drift. We introduce and discuss the average consensus protocol for frequency transfer in undirected networks where each node transmits and receives frequency information from other nodes. We analyze the following cases: 1) undirected networks with a static topology; 2) undirected networks with dynamic topology, where connections between nodes are made and lost dynamically; and 3) undirected networks with oscillator frequency drift. We show that all the nodes in a given network achieve average consensus and the number of iterations needed to achieve consensus can be minimized for a given cluster of nodes. Numerical simulations demonstrate that the consensus algorithm enables tolerable errors to obtain high coherent gain of greater that 90\% of the ideal gain in an error-free distributed phased array. 
\end{abstract}


\IEEEpeerreviewmaketitle


\section{Introduction}

\IEEEPARstart{C}{oherent} distributed arrays (CDAs) are clusters of cooperating wireless nodes whose radiated signals add up constructively at a designated destination. For instance, distributed beamforming~\cite{7803582,4202181}, virtual antenna arrays~\cite{1040381}, distributed MIMO arrays~\cite{4301290} are all implicit applications of CDAs. These distributed schemes have been proven to be technically feasible and bring improvements on three fronts:
\begin{enumerate}

\item \textit{Enhanced energy efficiency}: A transmit cluster focuses the emitted energy spatially toward the intended direction, yielding high directivity and thus better signal-to-noise ratio at the destination.

\item \textit{Increased reliability}: Having many transmitters allows for more signal path diversity and helps mitigate the multipath fading and shadowing.

\item \textit{Increased range}: Aggregate transmit power growth proportional to the square (and cube) of $N$ transmitters (i.e., radars) enables longer communication range.
\end{enumerate}

The major technical challenge in realizing these benefits consists of synchronizing the radio frequency (RF) signals on the array transmitters, so that they are coherently combined at the destination. This implies synchronization of the frequency and phase of these signals. In a distributed array, each RF signal is generated from an independent local oscillator. These oscillators have intrinsic frequency offsets and undergo stochastic frequency drifts over time (i.e., because of manufacturing tolerances and temperature variations) inducing phase offset variations. Therefore, it is necessary to compensate these clock offsets to frequency lock the transmitters and then implement phase calibration and channel estimation to achieve coherent transmission. CDAs generally require precise alignment of the phases of the transmitted signals to ensure energy-efficient transmissions to the destination~\cite{4202181}, with total phase errors of less than 18$^\circ$ required to achieve 90$\%$ ideal coherent gain, or a degradation of less than 0.5 dB from the ideal case \cite{7803582}.

Many techniques have been developed to achieve distributed transmitter synchronization, including receiver-coordinated explicit-feedback~\cite{6488994}, one-bit feedback~\cite{Mudumbai2006DistributedBU}, master-slave synchronization \cite{4202181}, reciprocity (i.e., channel state estimation by a transmit cluster from signals emitted by the target receiver)~\cite{bidigare2015wideband,7460546}, round-trip synchronization~\cite{4542555}, and two-way synchronization~\cite{5957340}. Most of these techniques use explicit/implicit feedback from the destination, and are thus \textit{closed loop CDAs}~\cite{7803582}. These feed-back schemes undergo latency and overhead, and most importantly are unsuitable for remote sensing and radar applications, where feedback is generally not available. 

In \cite{7803582}, the \textit{open loop CDA} (OL-CDA) concept was introduced, where the array self-aligns without the use of feedback from the destination.
Although this array uses zero feedback, it does involve inter-node coordination and phase alignment to achieve coherent beam steering. Indeed, it operates as distributed phased array, but the array geometry and element locations must be known. OL-CDA enables the development of future (fixed and mobile) wireless communication networks, which are energy efficient, secure, and can be built with low synchronization overheads. While closed loop CDA approaches have been heavily studied, the OL-CDA concept is still in its infancy, and more effort needs to be devoted to explore its potential. 
Recent theoretical studies and experiments have begun demonstrating the feasibility and potential applications of OL-CDAs (e.g. \cite{8058723,7839937,8168279}). OL-CDAs are inherently sparse antenna arrays, and while there has been considerable work done in the design and analysis of the radiation patterns of distributed antenna arrays in terms of the antenna locations and sidelobe levels (e.g. \cite{97353, 4148059, 6162964, 6193135, 6710134, 8072115, 8609181, 7588485}), there has not been a focused effort on the design of inter-node frequency alignment approaches  and the effects on distributed beamforming of frequency mismatch in the nodes in distributed antenna arrays in a decentralized sense.

A well-known frequency synchronization technique is the aforementioned master-slave architecture, where slave transmitters lock to a reference signal broadcast by the master transmitter. Despite its simplicity, this approach has the following limitations: (1) the technique is fragile to failures of the master node (if the master node fails, then the whole RF synchronization fails); (2) the architecture requires a dedicated master node, and thus limits the flexibility of the array. While GPS has been considered in some works as a master node, GPS is only adequate for coherent distributed transmissions at very low frequencies~\cite{5361473} ($\sim$10 MHz); (3) the master-slave architecture is not scalable.
	
This paper presents a new approach to distributed frequency alignment (syntonization) using decentralized (distributed) consensus averaging. Consensus, i.e., group agreement, has been considered in contexts such as flocking~\cite{1605401}, formation control~\cite{5427007},  flight of unmanned air vehicles (UAVs), and clusters of satellites. The proposed consensus scheme involves iterative frequency exchange (e.g., via local broadcast) among cooperating nodes to reach (asymptotically) a common agreed frequency, i.e., the average of all nodes’ initial frequencies. In other words, the frequency of each node converges to the global average. Note that unlike centralized master-slave schemes, the distributed scheme is robust to node failures and is scalable for large networks.

    This paper is organized as follows. Section II presents an analysis of the effect on coherent array gain of the phase error resulting from frequency misalignment. Section III models the network and introduces a distributed approach for frequency synchronization. For a static network topology, numerical results showing the effectiveness of the proposed algorithm are presented. Then Section IV employs the algorithm to deal with dynamic networks, where the connectivity of the network changes and the frequencies change due to oscillator drift. Numerical results show that a consensus can be reached in the presence of oscillator drift on low-cost oscillators to achieve above 90\% of the ideal level of coherent gain.

\section{Syntonization (Frequency Alignment) in Coherent Distributed Arrays}\label{sec:problem}

This work focuses on OL-CDAs, in which individual nodes self-align to achieve distributed coherence. As noted earlier, open-loop operation necessitates more information than closed-loop systems (where feedback from the destination is available). There are a few basic drivers of coordination errors in open-loop distributed arrays, which are errors in the measurement of the distances between node pairs; in aligning the phases of the nodes; in aligning the timing of the nodes; and in aligning the frequencies of the nodes. In previous work, we analyzed the effects on achieving coherent gain of errors in distance, phase, and time~\cite{7803582, 8378649}. In this work, we focus on the effects of errors in frequency alignment and present an approach for decentralized consensus optimization of the frequency. As such it is assumed that the errors due to distance, relative phasing, and timing are accounted for and thus negligible; because such errors are generally independent, their contributions can be analyzed individually and combined for a total error budget.

\begin{figure}[t!]
\centering
\begin{subfigure}[b]{0.24\textwidth}
   \includegraphics[width=1\linewidth]{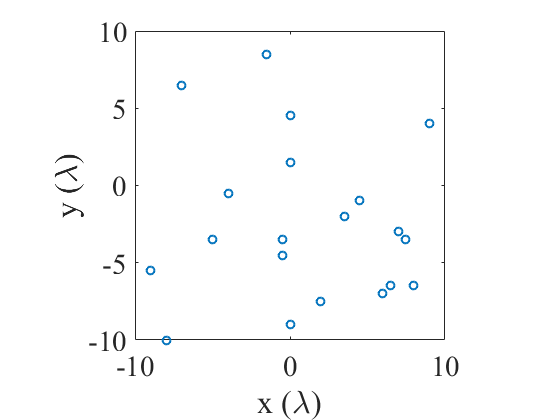}
  \caption{}
\end{subfigure}
\begin{subfigure}[b]{0.24\textwidth}
   \includegraphics[width=1\linewidth]{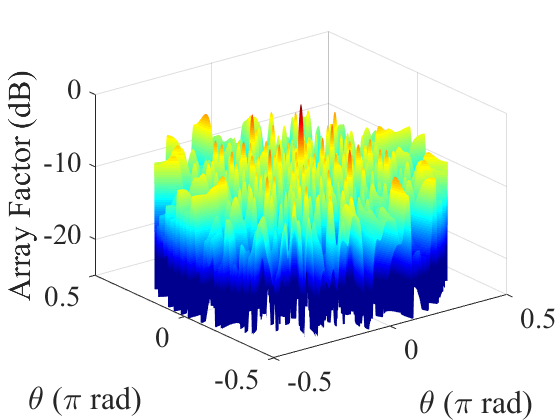}
 \caption{}
\end{subfigure}
 
\caption{(a) Sparse array layout for $N=20$ elements and (b) broadside radiation pattern.}\label{fig:N20}
\label{fig:array20} 
\end{figure}

\begin{figure}[t!]
\centering
\begin{subfigure}[b]{0.24\textwidth}
   \includegraphics[width=1\linewidth]{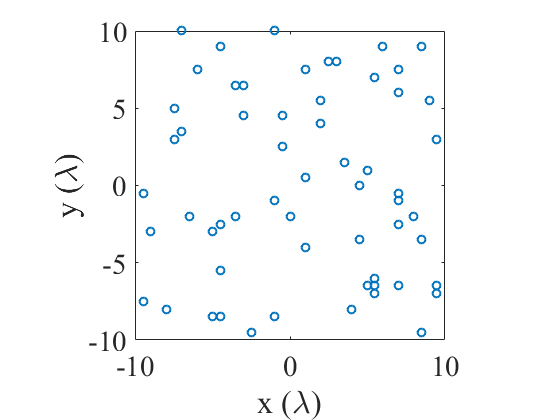}
  \caption{}
\end{subfigure}
\begin{subfigure}[b]{0.24\textwidth}
   \includegraphics[width=1\linewidth]{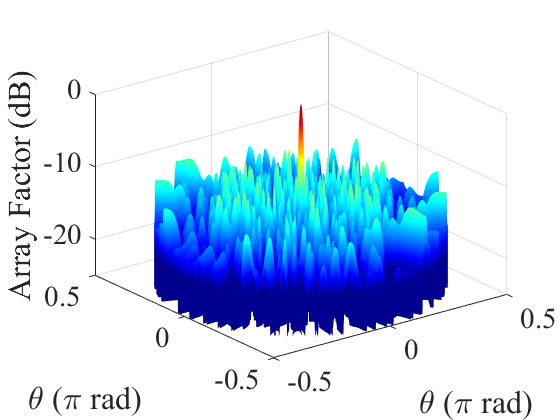}
  \caption{}
\end{subfigure}

\caption{(a) Sparse array layout for $N=60$ elements and (b) broadside radiation pattern.}\label{fig:N60}
\label{fig:array60}
\end{figure}

\begin{figure}[t!]
\centering
\begin{subfigure}[b]{0.24\textwidth}
   \includegraphics[width=1\linewidth]{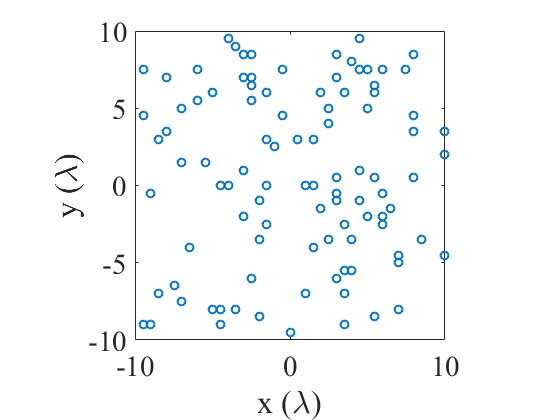}
  \caption{}
\end{subfigure}
\begin{subfigure}[b]{0.24\textwidth}
   \includegraphics[width=1\linewidth]{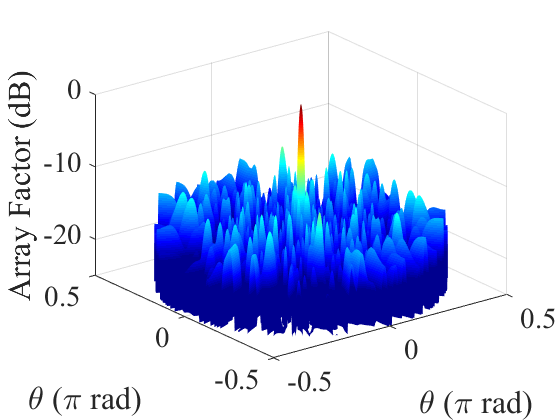}
  \caption{}
\end{subfigure}

\caption{(a) Sparse array layout for $N=100$ elements and (b) broadside radiation pattern.}\label{fig:N100}
\label{fig:array100} 
\end{figure}

In a coherent distributed transmit operation, the set of nodes in the array transmit waveforms with the appropriate relative phasing and timing such that the transmitted waveforms arrive at the target in-phase, ensuring that the waveforms add constructively at the destination node. We consider sparse planar arrays of transmitters, placed randomly within a $10\lambda\times10\lambda$ domain on a $\lambda/2$ grid. While the application space is general, notional use cases are communications from a set of compact radios (e.g. \cite{4202181}) or remote sensing from an array of UAVs (e.g. \cite{5172135, 8074868, 8072363, 8074924}); operating at a frequency of 100 MHz, the domain represents a 30m$\times$30m space, within which it is certainly feasible to locate nodes within half of a wavelength (1.5m). 
Figs.~\ref{fig:array20}-\ref{fig:array100} show random array layouts for arrays with 20, 60, and 100 nodes, and the corresponding array radiation patterns. The randomness in the array layout ensures that no significant sidelobes are present. When implementing frequency alignment between nodes, it is desired that the mainbeam gain remains as high as possible in the presence of errors. In the following we assess the degradation of the mainbeam gain in the presence of phase errors due to frequency misalignment.

\subsection{Distributed Array Model}

To model the relative effects of errors in a coherent distributed radar operation, we assess the relative coherent summation of the signals that is achieved in the presence of errors and  determine the resulting signal gain at a destination point relative to an ideal summation of signals without errors.
We consider the array to comprise vertically polarized small dipoles (antenna element factors could be included later in a straightforward manner), whose current density is 
\begin{equation}
\textbf{J} = \hat{\textbf{z}} e^{j\omega t}\sum_n I_n l_n \delta(x - dx_n) \delta(y - dy_n) \delta(z - dz_n),
\end{equation}
where $I_n$ is the current on element $n$, $l_n$ is the length of the $n^{th}$ dipole, and $(dx_n, dy_n, dz_n)$ is the displacement from the origin. The electric field intensity is found in terms of the far-field magnetic vector potential
\begin{equation}
\textbf{A} \approx \frac{e^{-jkr}}{4\pi r}\tilde{\textbf{J}},
\label{eq.vectorpotential}
\end{equation}
through
\begin{equation}
\textbf{E} = -j \omega \mu \textbf{A} + \frac{1}{j \omega \epsilon}\nabla\left(\nabla \cdot \textbf{A}\right),
\label{eq.e-field}
\end{equation}
where
\begin{align}
\tilde{\textbf{J}} &= \hat{\textbf{z}}e^{j\omega t}\sum_n I_n l_n e^{-j\left(k_x dx_n + k_y dy_n + k_z dz_n\right)}
\label{eq.current}
\end{align}
is the Fourier transform of the current density.

Beamsteering in a distributed array requires relative phase coordination between all elements in the array. Separate from aligning the frequency of the elements, phase coordination is what enables transmitted signals to arrive in-phase at the desired destination. The beamsteering phase of each element can be implemented in relative to a global array origin, or relative to other nodes in the array. When implementing a beamstearing phase to one element relative to the reference point, the wavenumber vector can be given by \cite{7803582}
\begin{equation}
\textbf{k}_n = \hat{\textbf{z}}_n\frac{2\pi}{\lambda}d_n\mathrm{cos}\theta_n 
\label{steering_phase}
\end{equation}
and the far-field electric field intensity is thus
\begin{align}
\mathrm{E}_\theta &=  jk\eta \mathrm{sin}\theta\sum_{n = 1}^{N} \frac{I_n l_n}{4 \pi r_n} e^{j\omega t}e^{-j\frac{2\pi}{\lambda}d_n \mathrm{cos}\theta_n} \nonumber \\
&= \sum_{n = 1}^{N} h_n \alpha_n e^{j\left(\omega t - \frac{2\pi}{\lambda}d_n \mathrm{cos}\theta_n\right)},
\end{align}
where $N$ is the number of elements in the array, $h_n$ contains the amplitude terms relating to the channel propagation effects and constant scalars, and $\alpha_n$ accounts for the amplitude of the current elements.
The signal $s(t)$ (V) received at a point in space is proportional to the electric field intensity E (V/m). If the receiving point is far-field to the array, and the array emits continuous-wave signals such that relative timing effects can be neglected, the total ideal signal can be written
\begin{equation}
s_i(t) = C\sum_{n = 1}^{N}{e^{j\left(2\pi ft - \frac{2\pi}{\lambda}d_n\mathrm{cos}\theta_n\right)}}.
\label{eq.ideal_gain}
\end{equation}
If errors are present in the coordination, the total received signal is given by
\begin{equation}
s_r(t) = C\sum_{n=1}^N{e^{j\left[2\pi(f+\delta f_n)(t - \delta t_n) + \phi_n + \phi_{s,n}\right]}},
\label{signal}
\end{equation}
where $\delta t_n$ is the time alignment error, $\delta f_n$ is the frequency error, $\phi_{s,n}$ is the phase added for beamsteering (from \eqref{steering_phase}), and $\phi_n$ is the phase error, given by
\begin{equation}
\phi_n = \frac{2\pi}{\lambda}\left(d_n + \delta d_n\right)\mathrm{cos}\left(\theta_n + \delta \theta_n\right) + \phi_c,
\end{equation}
where $\delta d_n$ is the error in the antenna separation measurement, $\delta \theta_n$ is the error in estimating the desired beamsteering angle relative to the platform attitude, and $\phi_c$ is the relative phase error of the oscillator.

\subsection{Signal Model with Frequency Errors}

While it is ultimately the differences in phase that cause degradation of the coherent gain, in this work the focus is on aligning the frequencies of the systems, or syntonization, between the separate platforms. In a system where continuous syntonization can be implemented, phase-locked loops (PLLs) can be used. In theory, the phase error of a PLL is zero, thus with continuous syntonization it can be expected that the phase error resulting from frequency differences will be negligible. In practice, however, syntonization may have to be implemented periodically, due to the nature of information transfer between the nodes. In this case, the frequencies of the oscillators on each node will drift between updates. The total phase error due to the evolution of the frequency drift in a time $T$ is given by
\begin{equation}
\delta\phi_{n} = 2\pi \delta f_n T.
\end{equation}
With other error terms corrected, the received signal in the presence of frequency drift is thus
\begin{equation}
s_r(t) = C\sum_{n=1}^N{e^{j\left(2\pi f t + \delta\phi_{n} + \phi_{s,n}\right)}}.
\label{freq_signal}
\end{equation}
If the frequency error is constant, the total phase error will be the frequency error multiplied by the time interval. In practice, the frequency will drift from zero error to some maximum value, thus estimating the frequency error as a constant represents an upper bound on the total error.

\subsection{Inter-Node Phase Error Tolerance Resulting From Frequency Drift}
To evaluate the effect of the error terms, the coherent gain $G_c$, which is the received signal power of \eqref{freq_signal} relative to (\ref{eq.ideal_gain}), and is given by
\begin{equation}
G_c = \frac{\left|s_r s_r^*\right|}{\left|s_i s_i^*\right|},
\label{Gc}
\end{equation}
is evaluated. The standard deviations of the frequency-induce phase error is varied, and the probability that the signal power exceeds a given threshold
\begin{equation}
P(G_c \geq X),
\end{equation}
where $0 \leq X \leq 1$ is the fraction of the ideal coherent signal power gain, is determined through Monte Carlo simulations. In this work, we evaluate the probability that the received signal will exceed 0.9 of the ideal gain ($X = 0.9$), corresponding to a coherent gain degradation of 0.5 dB.

\begin{figure}[t!]
\begin{center}
		\noindent
\includegraphics[width=3.2in]{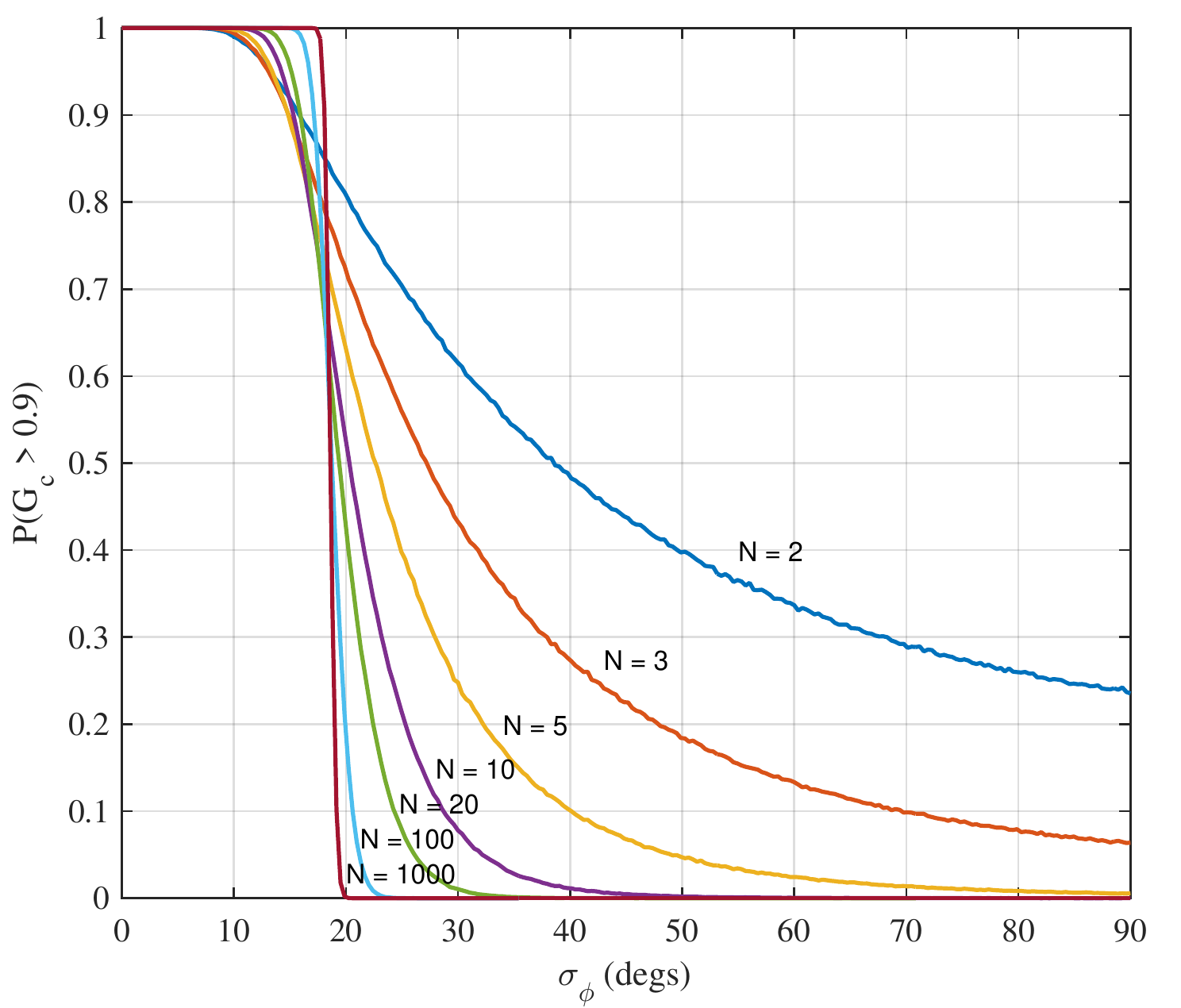}

\caption{Probability of the coherent gain exceeding $0.9$ versus clock phase error (100,000 Monte Carlo simulations). The threshold for achieving $P(G_c \geq 0.9) = 1$ approaches $18\degree$.}
\label{fig:prob}
\end{center}

\end{figure}

In Fig.~\ref{fig:prob}, the probability of the received signal power exceeding $0.9$ is shown for errors in the oscillator phase $\sigma_\phi$ for coherent distributed arrays consisting of $N$ = 2, 3, 5, 10, 20, 100, and 1000 nodes. As the number of nodes in the system increases, the line marking the area where $P(G_c \geq 0.9)$ becomes sharper, converging to a point below which $P(G_c \geq 0.9)$ is approximately 1, and above which is zero. The area where $P(G_c \geq 0.9) \approx 1$ also increases as the number of platforms increases. This is due to the decreasing variance of the received signal power as the number of elements in the network increases. As $N\rightarrow\infty$, this cutoff error below which $P(G_c \geq 0.9) \approx 1$ approaches $18\degree$. 

\begin{figure}[t!]
\centering
\begin{center}
\end{center}
\begin{subfigure}[b]{0.24\textwidth}
   \includegraphics[width=1\linewidth]{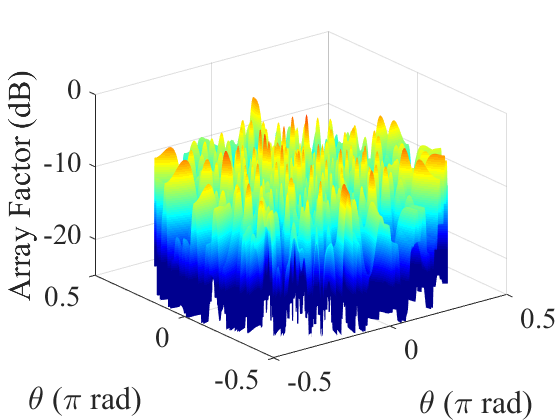}
  \caption{}
\end{subfigure}
\begin{subfigure}[b]{0.24\textwidth}
   \includegraphics[width=1\linewidth]{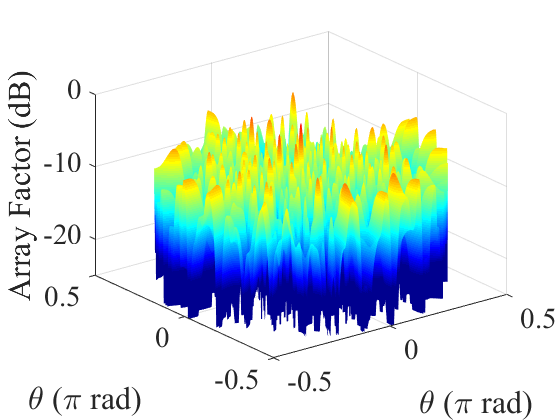}
  \caption{}
\end{subfigure}
\begin{subfigure}[b]{0.24\textwidth}
   \includegraphics[width=1\linewidth]{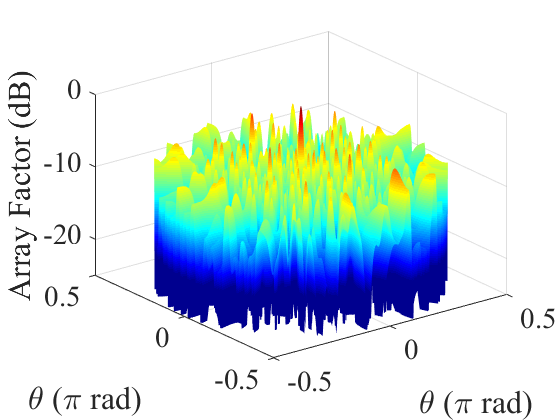}
  \caption{}
\end{subfigure}
\begin{subfigure}[b]{0.24\textwidth}
   \includegraphics[width=1\linewidth]{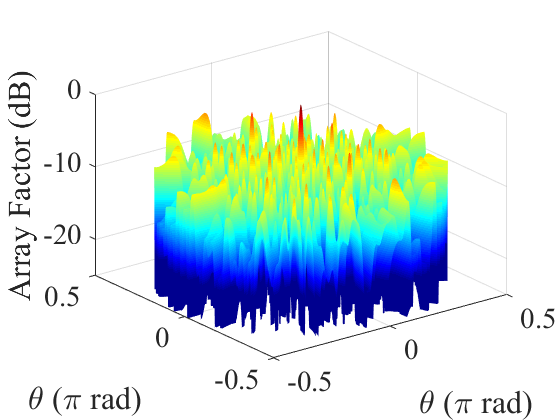}
  \caption{}
\end{subfigure}

\caption{Broadside beamforming for the $N=20$ element array with phase standard deviations of (a) $180\degree$, (b) $90\degree$, (c) $36\degree$, (d) $18\degree$.}\label{fig:N20ph}
\end{figure}

\begin{figure}[t!]
\centering
\begin{center}
\end{center}
\begin{subfigure}[b]{0.24\textwidth}
   \includegraphics[width=1\linewidth]{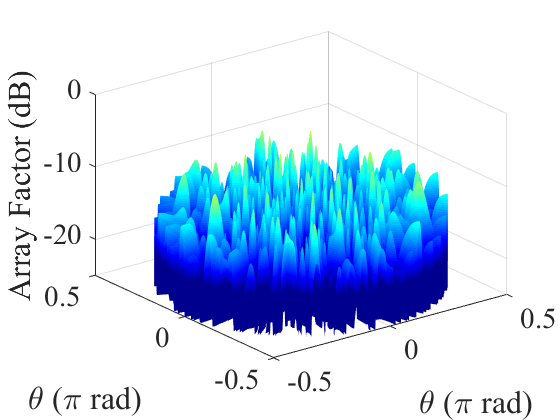}
  \caption{}
\end{subfigure}
\begin{subfigure}[b]{0.24\textwidth}
   \includegraphics[width=1\linewidth]{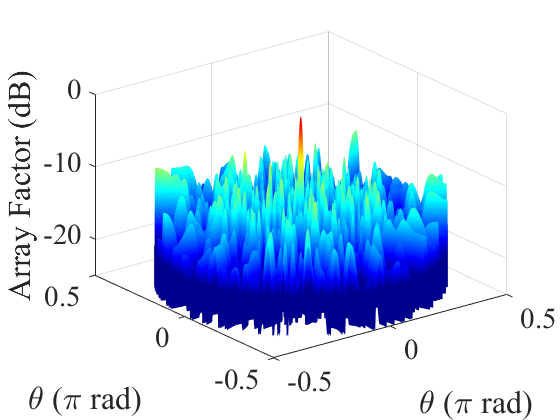}
  \caption{}
\end{subfigure}
\begin{subfigure}[b]{0.24\textwidth}
   \includegraphics[width=1\linewidth]{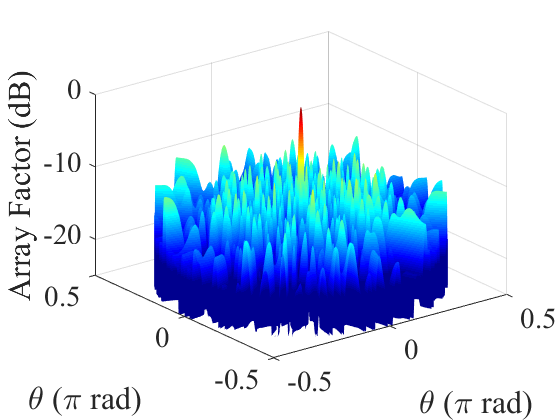}
  \caption{}
\end{subfigure}
\begin{subfigure}[b]{0.24\textwidth}
   \includegraphics[width=1\linewidth]{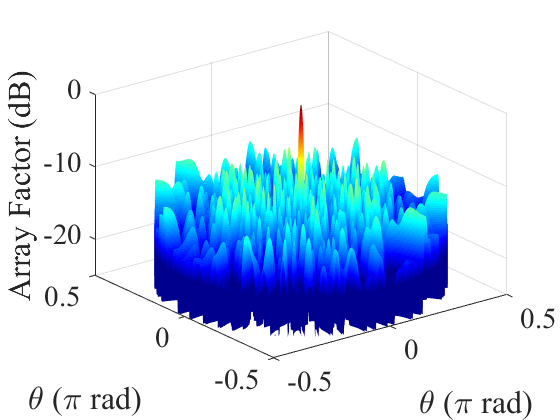}
  \caption{}
\end{subfigure}

\caption{Broadside beamforming for a $N=60$ element array with phase standard deviations of (a) $180\degree$, (b) $90\degree$, (c) $36\degree$, (d) $18\degree$.}\label{fig:N60ph}
\end{figure}

\begin{figure}[t!]
\centering
\begin{center}
\end{center}
\begin{subfigure}[b]{0.24\textwidth}
   \includegraphics[width=1\linewidth]{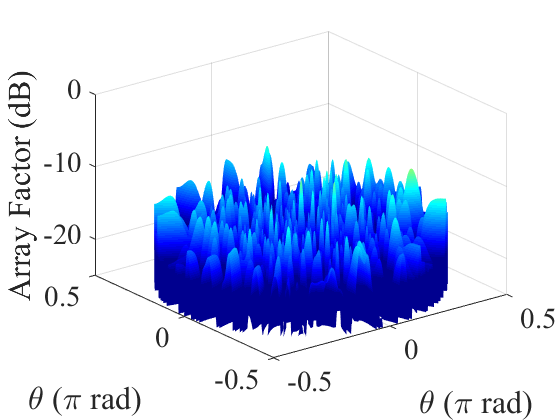}
  \caption{}
\end{subfigure}
\begin{subfigure}[b]{0.24\textwidth}
   \includegraphics[width=1\linewidth]{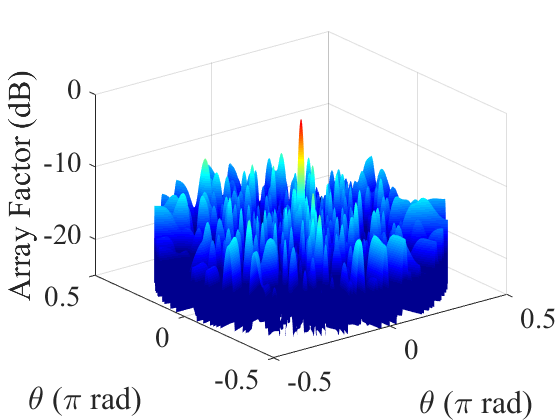}
  \caption{}
\end{subfigure}
\begin{subfigure}[b]{0.24\textwidth}
   \includegraphics[width=1\linewidth]{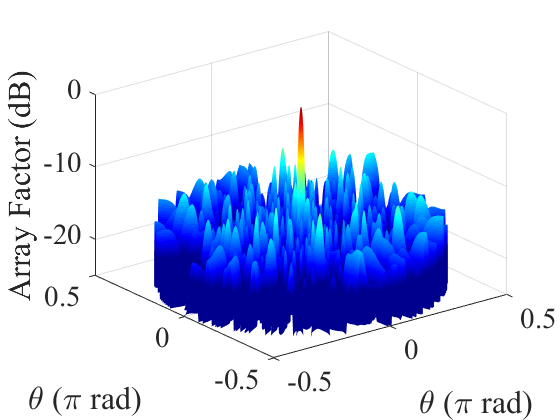}
  \caption{}
\end{subfigure}
\begin{subfigure}[b]{0.24\textwidth}
   \includegraphics[width=1\linewidth]{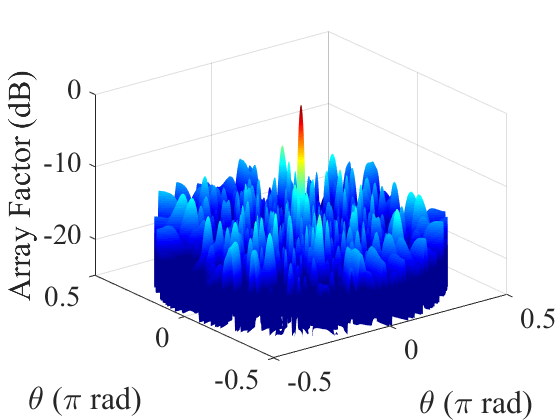}
  \caption{}
\end{subfigure}

\caption{Broadside beamforming for the $N=100$ element array with phase standard deviations of (a) $180\degree$, (b) $90\degree$, (c) $36\degree$, (d) $18\degree$.}\label{fig:N100ph}
\end{figure}

The effect of phase errors on the array beampattern is shown in Figs.~\ref{fig:N20ph}-\ref{fig:N100ph}. For relatively small arrays ($N=20$), the mainbeam degradation relative to the sidelobe levels is significant when the phase error is above $36\degree$, and the mainbeam gain is in the order of the sidelobe gain. At lower errors, the mainbeam increases, and at $18\degree$ the mainbeam gain is degraded by only 0.5 dB. The sidelobe levels are low for larger arrays of $N=60$ and $N=100$, thus the beam can still be formed with errors up to $90\degree$, albeit with significant gain degradation. 

\section{Decentralized Frequency Alignment}\label{sec:algorithm}
\subsection{System Model}
The distributed antenna array is modeled as a network (connected undirected graph) $G=\{\mathcal{N},\mathcal{E}\}$ consisting of a set of nodes $\mathcal{N}=\{1,\cdots,n\}$ and a set of undirected edges $\mathcal{E}$. 
An edge $(i,j) \in \mathcal{E}$ means that there is a connection between nodes $i$ and $j$ and the communication is bidirectional. 
We assume that the nodes are equipped with hardware clocks subject to clock drift.

Each node $i$ has an associated RF signal ${x}_i(t)=\cos(2\pi f_i t)$, the frequency of which the network must reach a consensus.  
It has an initial estimate for the true frequency $f_c$, which is denoted as $f_i(0)$. 
Then decentralized averaging algorithms update the estimates at all nodes during each iteration. 
We say that the nodes reach a consensus if 
\begin{equation}
f_i=\bar f = {1\over n}\sum_{i=1}^n f_i(0),\quad \forall i=1,\dots,n.
\end{equation}
We assume that $f_i(0) = f_c + \sigma_iX$ where $\sigma_i = 1\times10^{-4} f_c$ is the error in frequency: $1\times 10^{-4}$ i.e., $100$ ppm (parts per millions) represents crystal clock accuracy, and $X$ is a random variable with normal distribution $N(0,1)$.

\subsection{Average Consensus Model}
		
The proposed decentralized frequency alignment consensus (DFAC) scheme with a fixed graph is described in Algorithm~\ref{alg:1}. 
We denote $\mathbf{f}(k)=\{f_1(k),\cdots,f_n(k)\}$ as the estimates from all nodes at iteration $k$.

\begin{algorithm}
\caption{Decentralized frequency alignment consensus}\label{alg:1}
\begin{algorithmic}
\State {\bf Input:}  $k=0$, the initial estimates $\mathbf{f}(0)$, mixing matrix $\mathbf{W}$
\While{stopping criteria is not satisfied}
    \State $k = k+1$
    \State $\mathbf{f}(k)= \mathbf{W}\mathbf{f}(k-1)$ 
\EndWhile

\State \Return $\mathbf{f}(k)$
\end{algorithmic}
\end{algorithm}

The mixing matrix $\mathbf{W} = [{w}_{ij}]$, where ${w}_{ij}$ is the $(i,j)$ element of the matrix, satisfies the following assumptions:
\begin{itemize}
    \item (Symmetry) $\mathbf{W} = \mathbf{W}^\top$.
    \item (Doubly stochastic) $\mathbf{W}\textbf{1} = \textbf{1}$,  $\textbf{1}^\top \mathbf{W}= \textbf{1}^\top$, where $\mathbf{1}$ is the $n\times 1$ vector of all ones. That is, each row sums to 1 and each column sums to 1.
    \item (Decentralized property) ${w}_{ij} = 0$ if $i\neq j$ and nodes $i$ and $j$ are not connected, i.e., $(i,j) \notin \mathcal{E}$. 
\end{itemize}

\subsection{Numerical Results}
In this section, we evaluate the performance of the proposed DFAC algorithm in a simulated environment with a fixed network. The algorithm is implemented in MATLAB; the initial states of the nodes are selected at $f_c$ = 1 GHz. We evaluate the quantitative performance of the decentralized algorithm by computing the normalized error $( \textbf{f}-{f}_c)/ ({f}_c)$ along the iteration number. 
We vary the number of nodes $N$ and the network connectivity.

\subsubsection{Varying number of nodes}

\begin{figure}[t!]
\begin{center}
		\noindent
\includegraphics[width=3.5in]{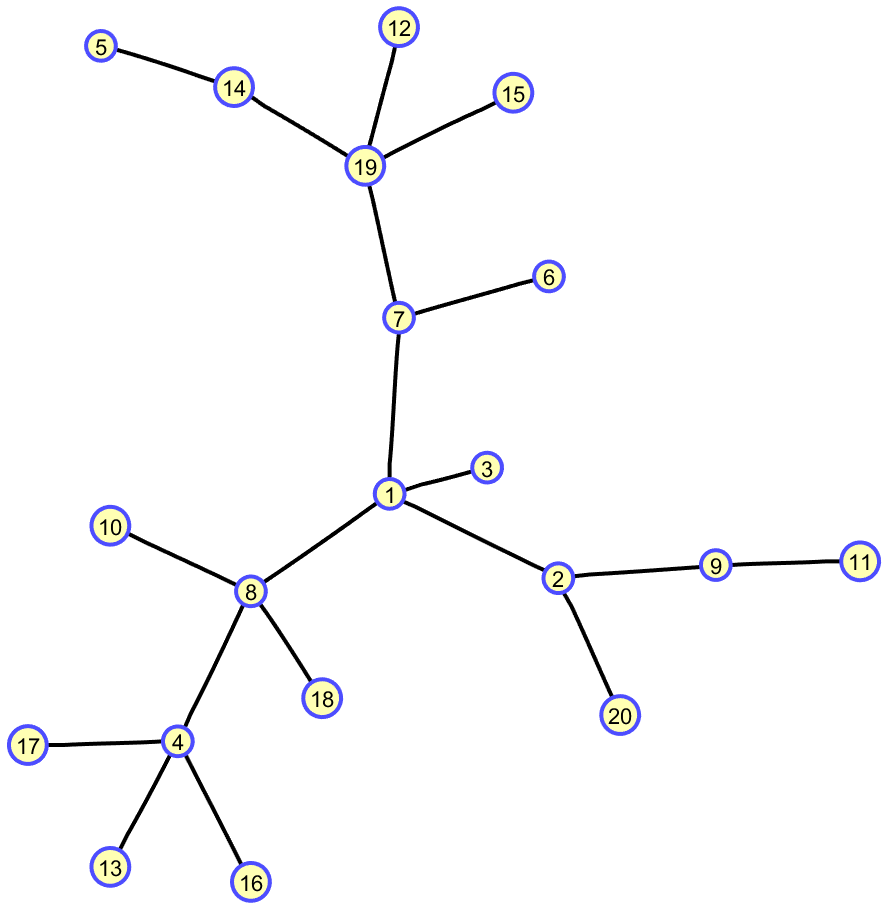}

\caption{A sparsely connected network of 20 nodes and $r=0.1$.}
\label{fig:sparce}
\end{center}

\end{figure}

\begin{figure}[t!]
\begin{center}
	\noindent
\includegraphics[width=3.6in]{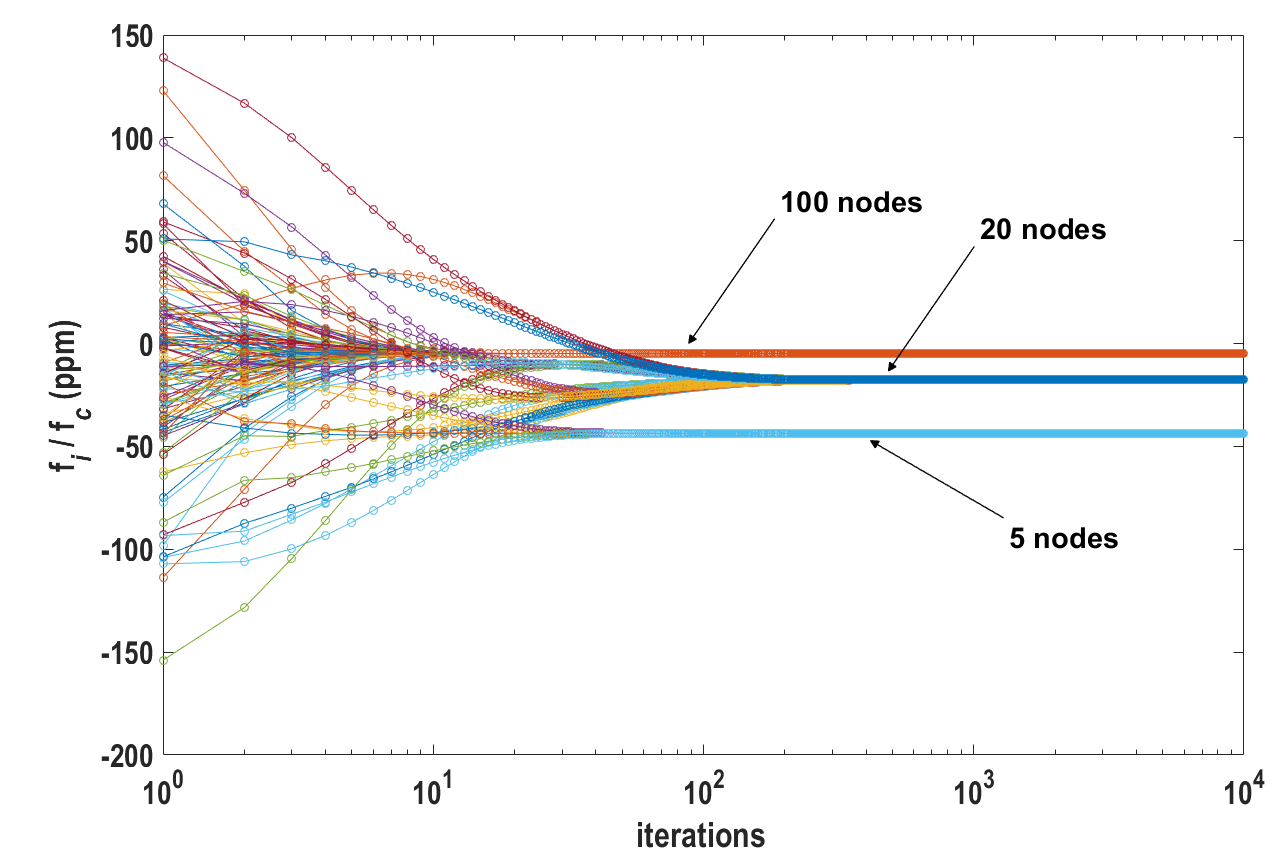}
\caption{Normalized error comparison under varying number of nodes of sparse networks with $r = 0.1$ (20 and 100 nodes) and $r = 0.4$ (5 nodes).}
\label{fig:net}
\end{center}

\end{figure}

The network is randomly generated with connectivity ratio $r$, where $r$ is defined as the number of edges divided by the total number of all possible edges $N(N-1)/2$. When $r$ is small, the network presents a cluster of sparsely connected nodes. As $r$ increases, the network becomes more and more connected. 
Fig.~\ref{fig:sparce} is a sparse network of 20 nodes. 
The total number of possible edges is 190, and this network has 19 edges.
We fix $r=0.1$ and change the number of nodes $n$ to 5, 20, and 100.
The result is shown in Fig.~\ref{fig:net}. 
This figure shows that as $n$ increases, the final absolute value of error decreases. 
In addition, the algorithm converges faster with more nodes (100 nodes and 20 nodes). It is worth noting that the $r$ value for 5 nodes is 0.4 (the minimum number of edges is 4) resulting in faster convergence than 20 nodes. 

\begin{figure}[!ht]
\begin{center}
		\noindent
\includegraphics[width=3.6in]{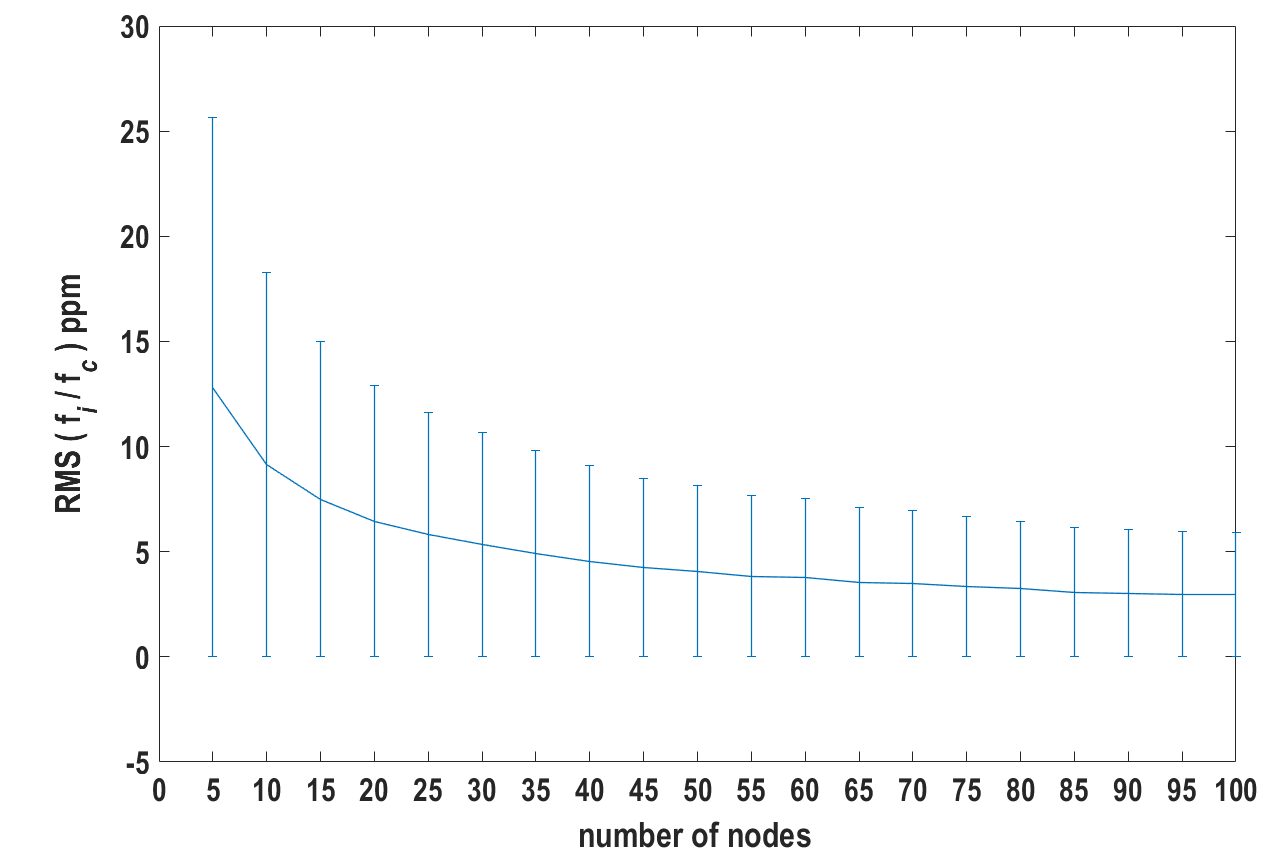}

\caption{RMS error for varying networks.}
\label{fig:RMS}
\end{center}

\end{figure}

Fig.~\ref{fig:RMS} shows the RMS Error over 10,000 simulations for different numbers of nodes $n$ between 5 and 100. i.e., we averaged the normalized error at all 10,000 simulations for all $n$ values. 
The case $n$ = 5 gives the highest error with the largest variability. As the number of nodes increases, the error decreases and its variability lessens. Large values of $N$ achieve similar performance; in particular, they have small errors. 
Indeed, the RMS Error for a small size network with 5 nodes is 12.8 ppm, it reaches about 2.9 ppm and 4.1 ppm for large networks with 100 and 60 nodes, respectively, while 6.5 ppm error is obtained by a medium size network (20 nodes). Thus the large 60 and 100 nodes networks save  $43\%$ and $55\%$ of the error, respectively comparing to a 20 node network.

It is important to note that the error holds irrespective of the connectivity $r$ of the network, as will be demonstrated later.

\begin{figure}[!t]
\begin{center}
		\noindent
\includegraphics[width=3.6in]{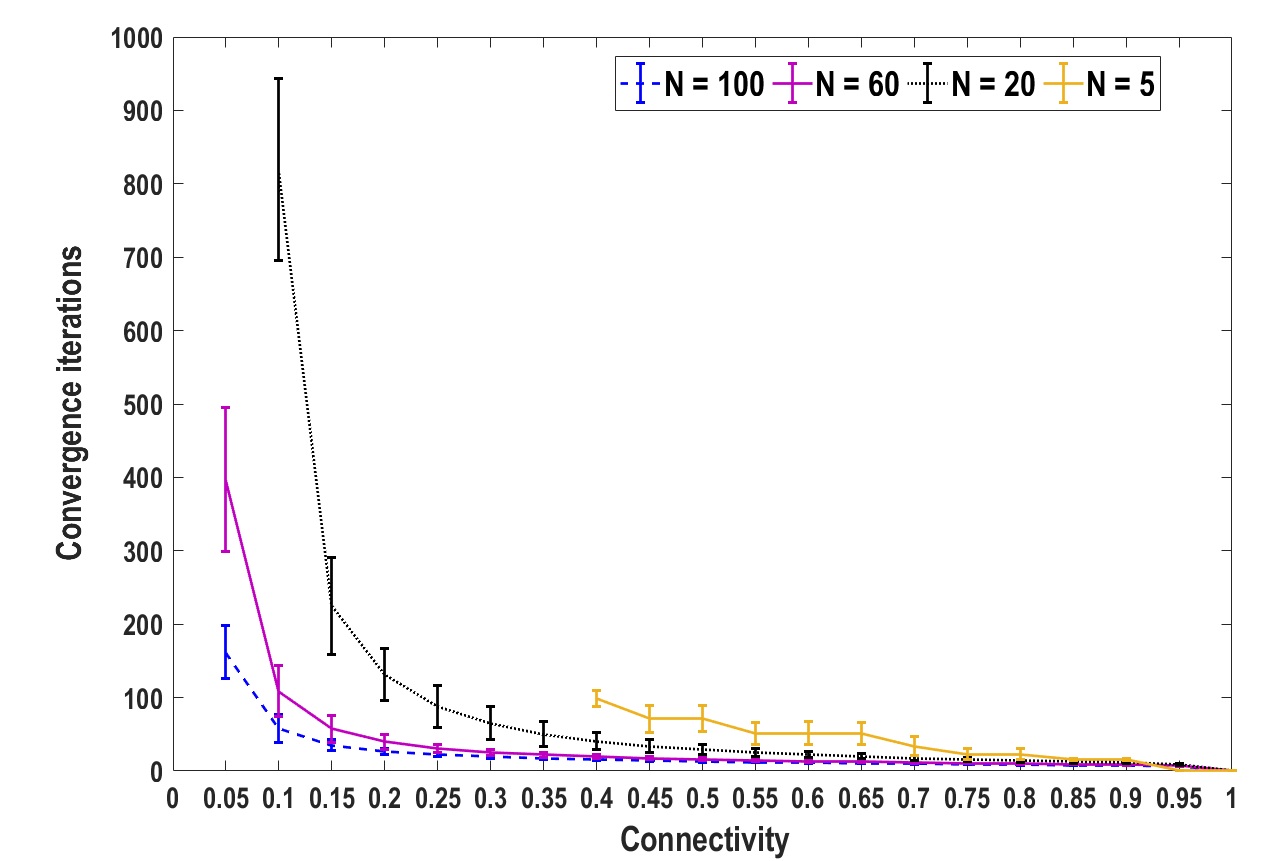}

\caption{Convergence iterations for varying networks.}
\label{fig:CS}
\end{center}
\end{figure}

Then we compare the convergence speed for different connectivity. 
We measure the number of iterations required for consensus. We evaluate the stopping criteria for DFAC algorithm as $\max|\mathbf{f}(k)-\bar f| < 2\times 10^{-3}$. The results for $n=5,~20,~60$, and~$100$ are shown in Fig.~\ref{fig:CS}. 
We plotted the means and variances for 10,000 simulations. 
For all values of $r \geq 0.1$, the two large  clusters (60 and 100 nodes) perform better than the other two (5 and 20 nodes). For example, with $r = 0.1$, the number of iterations for 100 nodes is about 57. It reaches 819 for 20 nodes, while the second smallest number is attained by 60 nodes that requires around 108 iterations. Hence the 100 nodes network saves about 47$\%$ of the iterations comparing to 60 nodes, and this induces 47$\%$ less transmissions among the nodes. Though the number of iterations is smaller for 100 nodes, the number of communications is larger for each iteration, compared to 60 nodes. This is because the number of edges is bigger (on average there are 6 nodes connected to one node for a 60 nodes network, while the number increases to 10 for a 100 node network). As the connectivity $r$ increases, the number of iterations decreases for all networks. A large connectivity (high density of links) is probably unrealistic as it would promote interference among the nodes.
In fact, reducing the convergence time is important, as it leads to a smaller number of messages exchanged among the nodes and to a reduced energy cost of the algorithm.

\subsubsection{Varying connectivity}

\begin{figure}[t!]
\centering
\begin{center}
\end{center}
\begin{subfigure}[b]{0.5\textwidth}
   \includegraphics[width=1\linewidth]{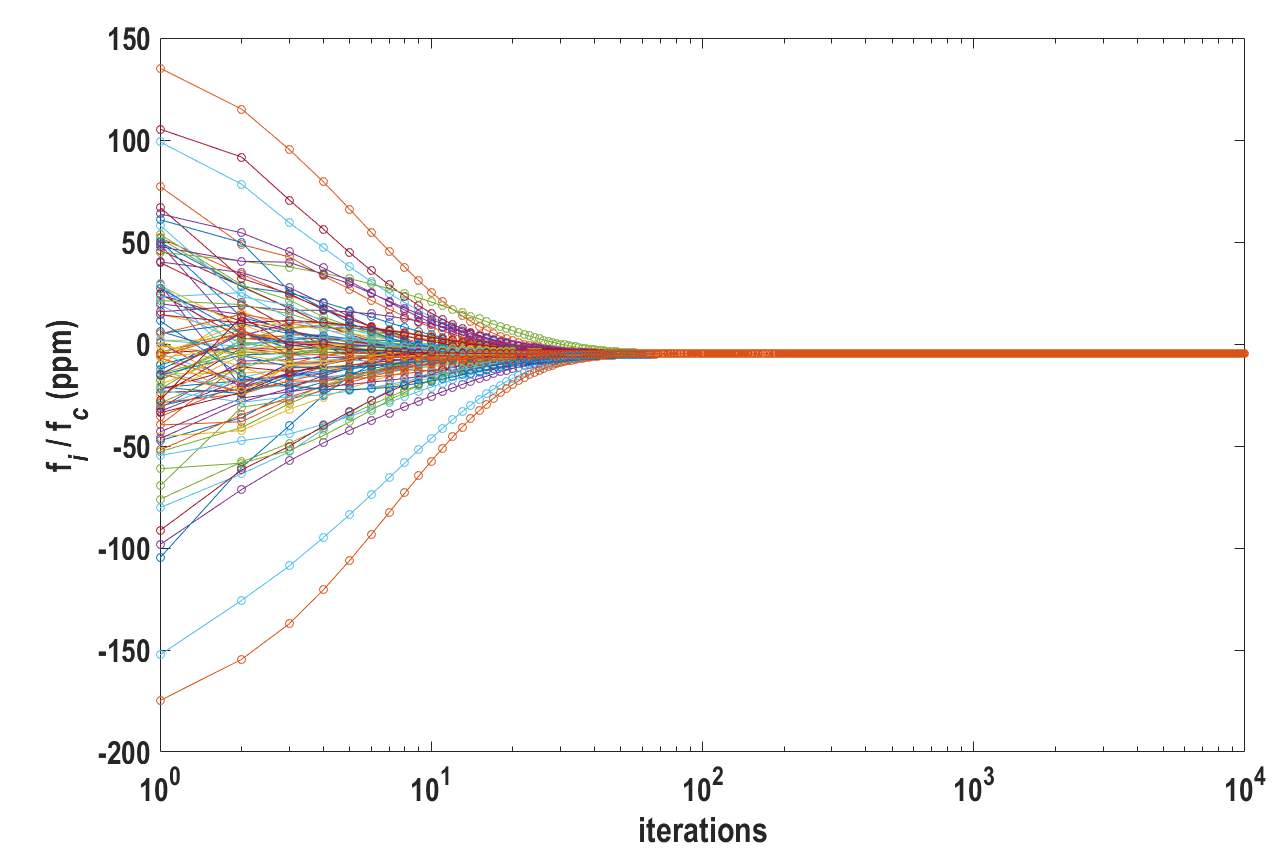}
  \caption{}
\end{subfigure}

\begin{subfigure}[b]{0.5\textwidth}
  \includegraphics[width=1\linewidth]{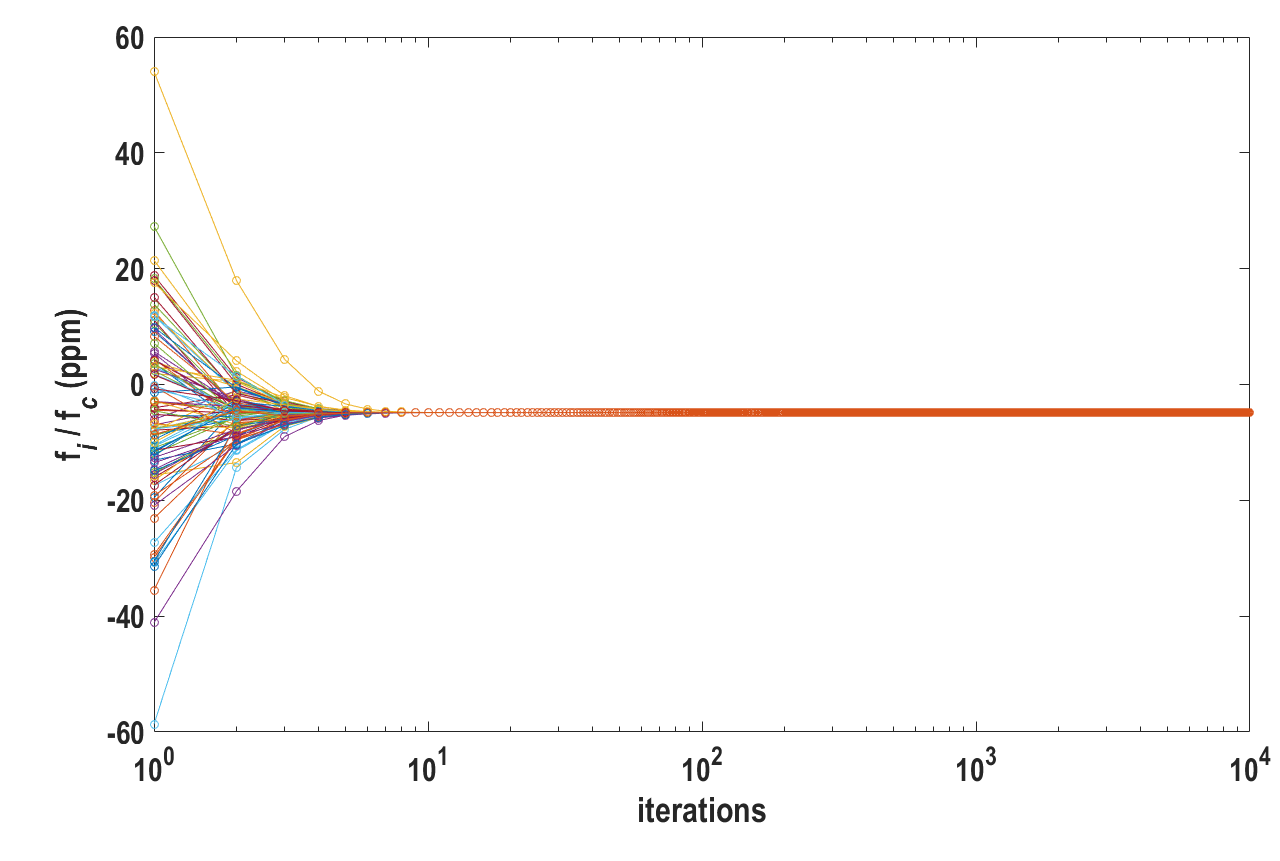}
   \caption{}
\end{subfigure}

\begin{subfigure}[b]{0.5\textwidth}
  \includegraphics[width=1\linewidth]{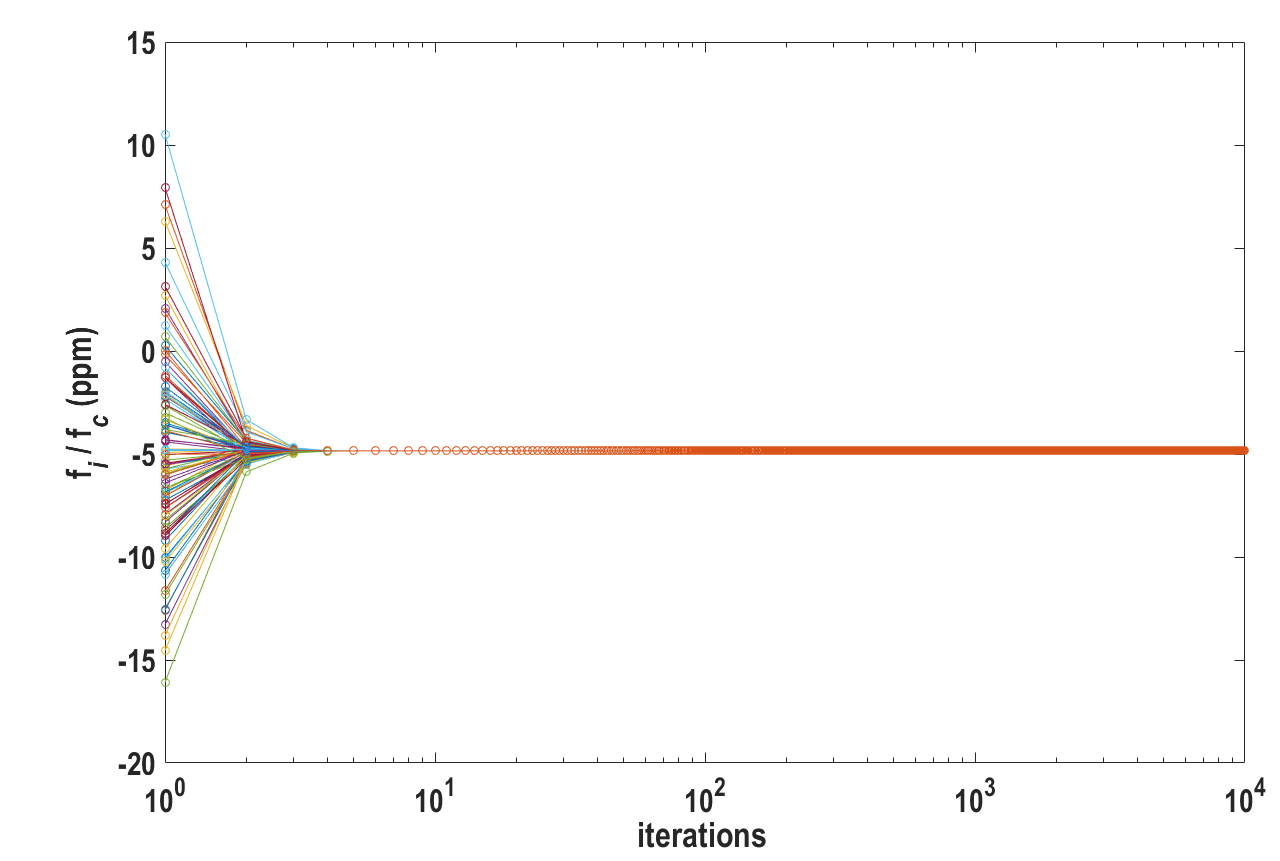}
   \caption{}
\end{subfigure}

\caption[simulations]{State trajectories of all nodes corresponding to network with $n$ = 100. (a) $r=0.05$. (b) $r=0.3$. (c) $r=0.9$.}\label{fig:ST}
\end{figure}

For $n$ = 100, the sparsity of the network is changed from a sparse connectivity ($r = 0.05$) to a dense connectivity ($r = 0.9$). The state trajectories are shown in Fig.~\ref{fig:ST}. It is clear that as the number of edges increases, connectivity increases, and the settling time of the state trajectories decreases. 
An explanation of this trend can be inferred by looking at $\lambda_2(\mathbf{W})$ (the second largest eigenvalue in magnitude) of the undirected graph that is tightly linked to the convergence speed of the algorithm. Indeed, $\lambda_2$ is relatively small for dense networks and is relatively large for sparse networks. Therefore, a dense network solves a consensus problem faster than a sparse network.

It can be seen that the error remains unchanged for changing connectivity $r$. This is solely due to the rational of the average consensus, i.e., the average consensus algorithm guarantees that the iterative procedure $\mathbf{f}(k) = \mathbf{W} \mathbf{f}(k-1)$ leads all the estimates to converge to ${\bar{f}}$, irrespective of the connectivity of the network as long as this network is connected.

\section{DYNAMIC FREQUENCY ALIGNMENT IN COHERENT DISTRIBUTED ARRAYS}

To investigate the performance of the average consensus algorithm in a realistic environment, we consider two scenarios: changing mixing matrix and changing frequencies due to oscillator drift.  

\subsection{Dynamic Array Connectivity}

In Section~\ref{sec:algorithm}, the network topology is assumed stationary. But wireless networks are intrinsically dynamic, i.e., due to the presence of obstructions, or fluctuating detection range.  In this regard, the effects on the consensus of a dynamic network, i.e., a network with a time-varying topology, should be taken into consideration. 
Herein we are interested to investigate the convergence time of a dynamic network. We assume a varying interconnection topology in which the number of nodes is fixed but the communication links are changing; links are restored and failed. i.e., edges are added and removed from the graph. Algorithm~\ref{alg:2} describes the detailed operation of changing the mixing matrix.

\begin{algorithm}
\caption{DFAC with dynamic networks}\label{alg:2}
\begin{algorithmic}
\State {\bf Input:}  $k=0$, the initial estimates $\mathbf{f}(0)$, mixing matrix $\mathbf{W}$, probabilities $P_1\leq 1$
\While{stopping criteria is not satisfied}
    \State $k = k+1$
    \State Generate a random number in $\alpha\in [0,1]$ from the uniform distribution 
    \If{$\alpha\in (P_1,P_1+(1-P_1)/2]$} 
    \State Randomly choose one node $i$
    \State Randomly choose another node $j$ in its neighbor 
    \State Change the submatrix of $\mathbf{W}$ with $i,j$ rows and columns: $[w_{ii},w_{ij},w_{ji},w_{jj}]\rightarrow [w_{ii}+w_{ij},0,0,w_{ji}+w_{jj}]$
    \ElsIf {$\alpha > P_1+(1-P_1)/2]$}
    \State Randomly choose one node $i$
    \State Randomly choose another not connected node $j$ 
    \State Change the submatrix of $\mathbf{W}$ with $i,j$ rows and columns: $[w_{ii},0,0,w_{jj}]\rightarrow [w_{ii}-w,w,w,w_{jj}-w]$, with $w=0.2\times\min\{w_{ii},w_{jj}\}$
    \EndIf
    \State $\mathbf{f}(k)= \mathbf{W}\mathbf{f}(k-1)$ 
\EndWhile

\State \Return $\mathbf{f}(k)$
\end{algorithmic}
\end{algorithm}

\begin{figure}[t!]
\centering
\begin{center}
\end{center}
\begin{subfigure}[a]{0.5\textwidth}
   \includegraphics[width=1\linewidth]{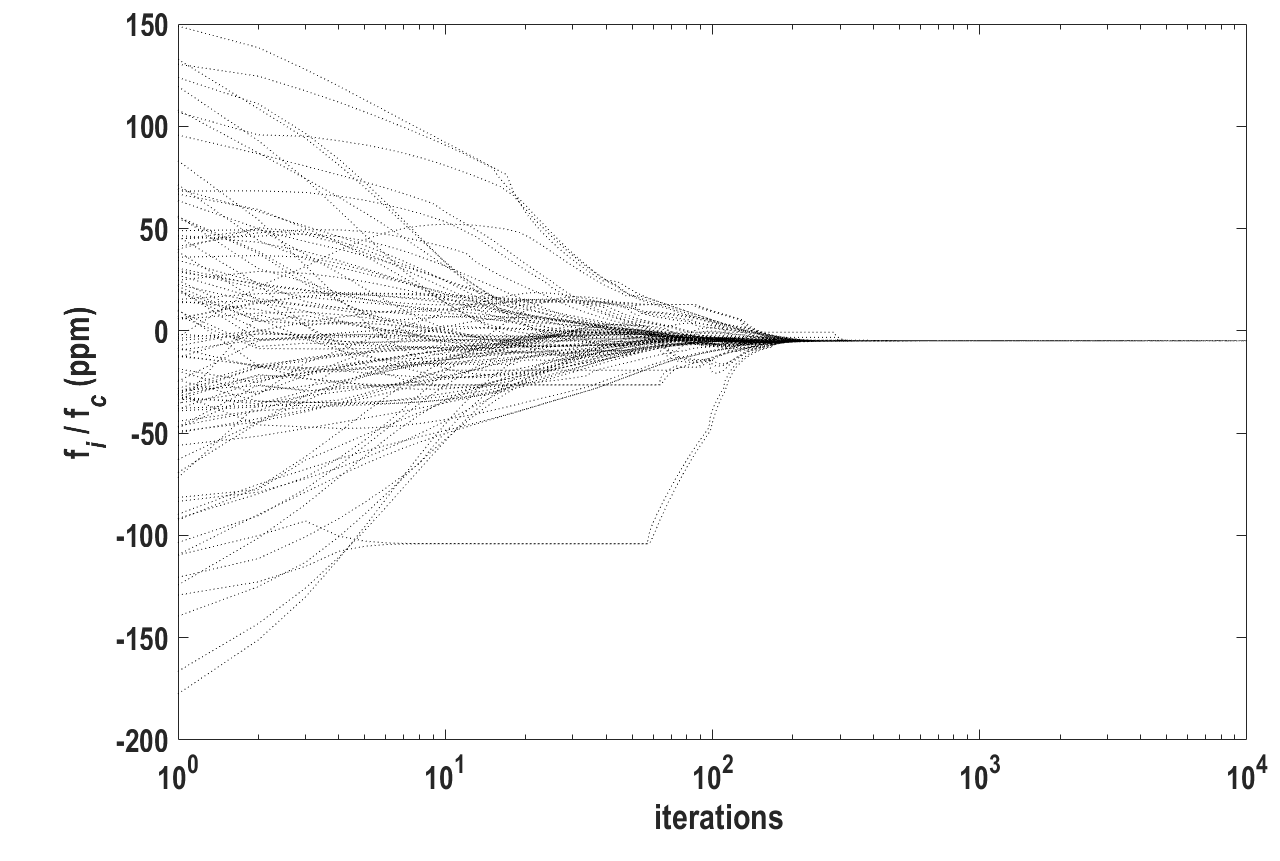}
  \caption{}
\end{subfigure}

\begin{subfigure}[b]{0.5\textwidth}
  \includegraphics[width=1\linewidth]{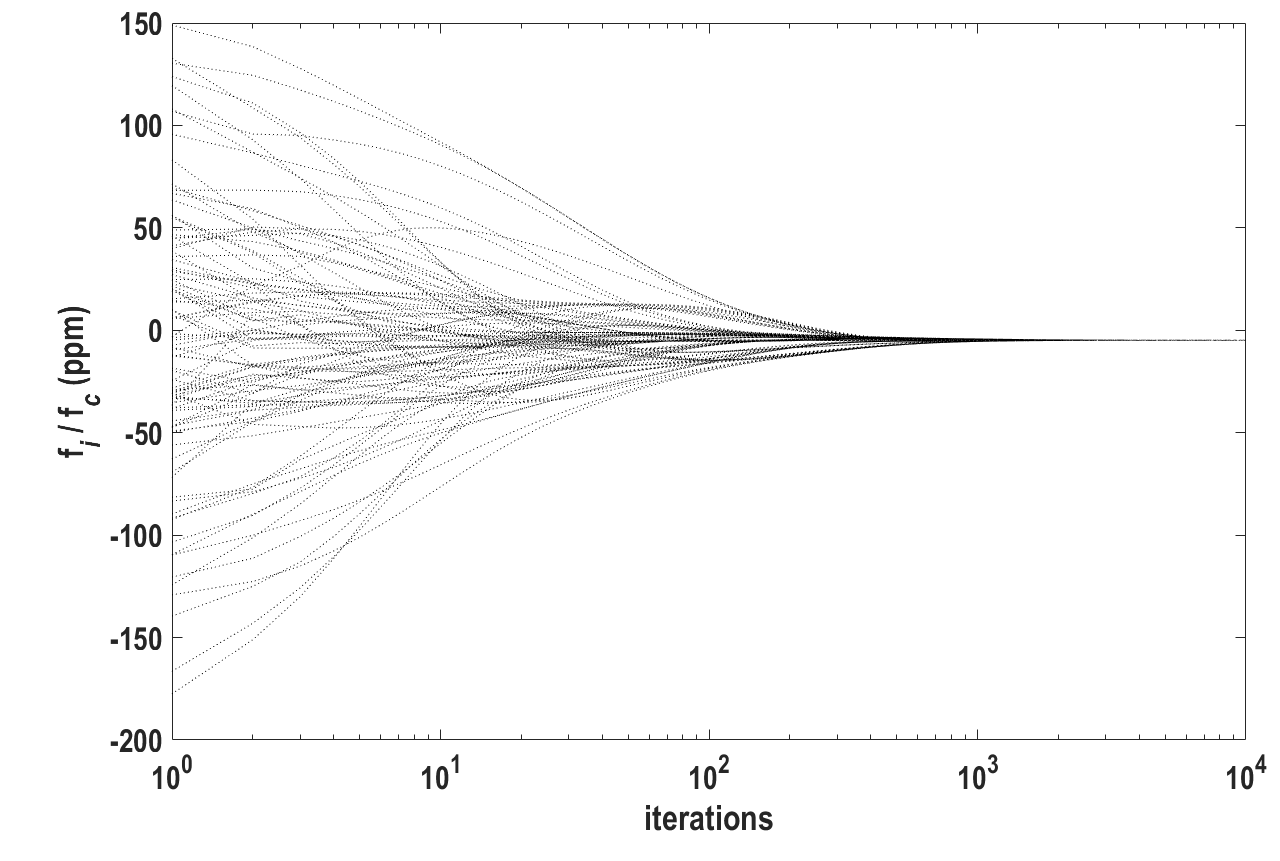}
   \caption{}
\end{subfigure}

\caption[Two simulations]{State trajectories of all nodes corresponding to network with $N$ = 100 with $r=0.03$. (a) Dynamic network. (b) Static network.}\label{fig:DNST}
\end{figure}

\begin{figure}[t!]
\begin{center}
\includegraphics[width=0.51\textwidth]{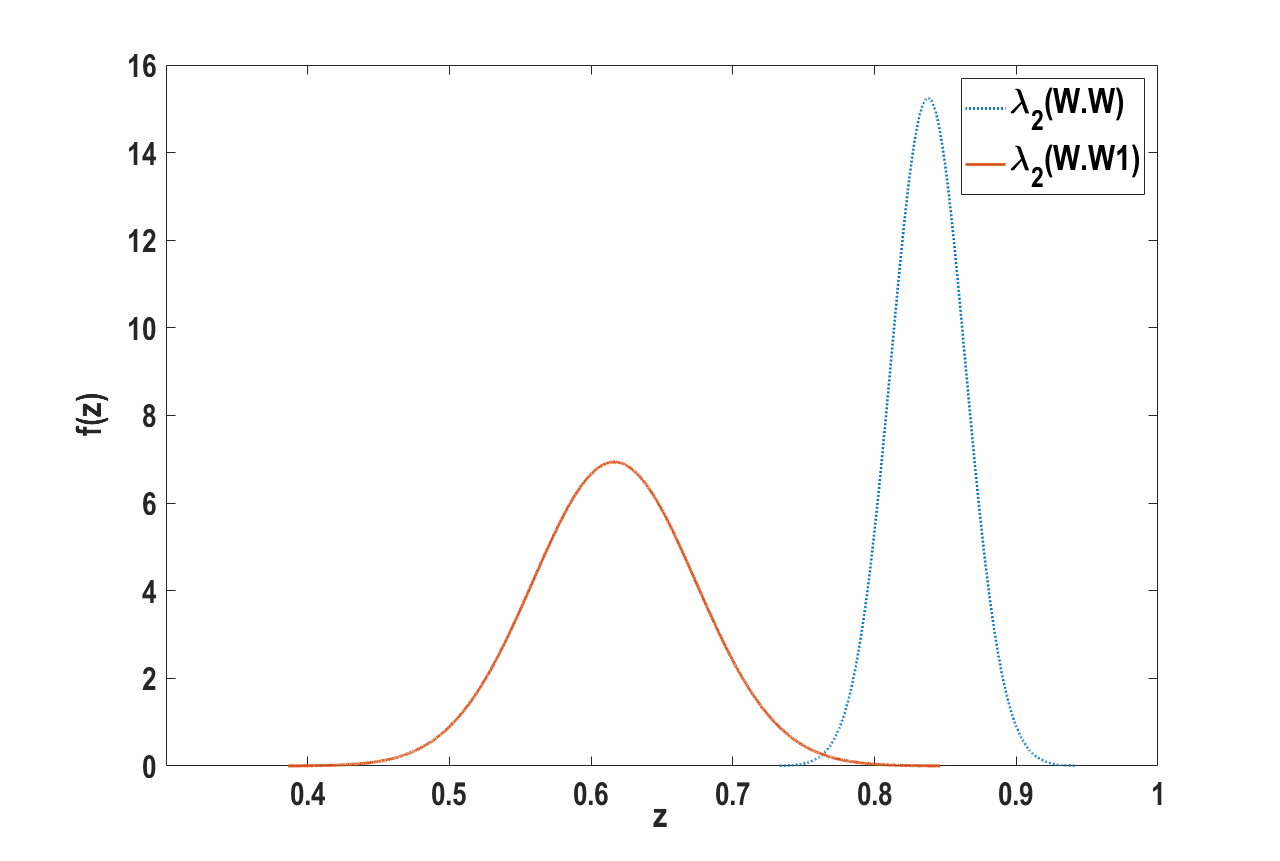}
\caption{Normal distribution of the second largest eigenvalue $\lambda_2$ samples of changing and unchanging mixing matrices.
$N=100, r=0.05$.}
\label{fig:ND}
\end{center}
\end{figure}

We consider a network with $n$ = 100 nodes and connectivity $r=0.03$. 
At every iteration $k$, the topology of the network is reconfigured. 
A number $\alpha$ is randomly generated from a uniform distribution from $[0,1]$.
If $\alpha<1$, then we do not change the mixing matrix. Otherwise, depending on the value of $\alpha$, we either add or remove one edge from the network. 
When we remove one edge $(i,j)$, we move the weight $w_{ij}$ and $w_{ji}$ onto the nodes $i$ and $j$. That is $[w_{ii},w_{ij},w_{ji},w_{jj}]\rightarrow [w_{ii}+w_{ij},0,0,w_{ji}+w_{jj}]$.
When we add a new edge $(i,j)$, we move the weight $w$ from nodes $i$ and $j$ to the edge, i.e., $[w_{ii},0,0,w_{jj}]\rightarrow [w_{ii}-w,w,w,w_{jj}-w]$, with $w=0.2\times\min\{w_{ii},w_{jj}\}$.
These operations keep the properties of the mixing matrix. 
When $P_1 = 1$, no changes occur and the matrix is not reconfigured. 
Fig.~\ref{fig:DNST}(a) shows the state trajectories of this dynamic network. 
One can observe that a consensus is asymptotically reached. The results of the original case (static network with $n$ = 100 and $r=0.03$ are shown in Fig.~\ref{fig:DNST}(b) for comparison). 
As can be seen from both graphs, the reconfiguration decreases the settling time of the state trajectories. 
This character is consistent with the above results that explain the decrease in convergence time by the relatively unpronounced eigenvalue $\lambda_2$ of the network. 
Assume that the matrix $W1$ is changed to $W$. 
Then $\lambda_2(W\cdot W1)$ of the product of changing mixing matrix from $W1$ to $W$ is consistently less than $\lambda_2(W\cdot W)$ of the product of the matrix by itself (no changes in the matrix). 
To support this argument, Fig.~\ref{fig:ND} compares the normal distribution curves of 10,000 sampled $\lambda_2$ of changing and unchanging mixing matrices. It can be seen that $\lambda_2(W\cdot W1)<\lambda_2(W\cdot W)$ with a high probability.

\begin{figure}[t!]
\begin{center}
\includegraphics[width=0.5\textwidth]{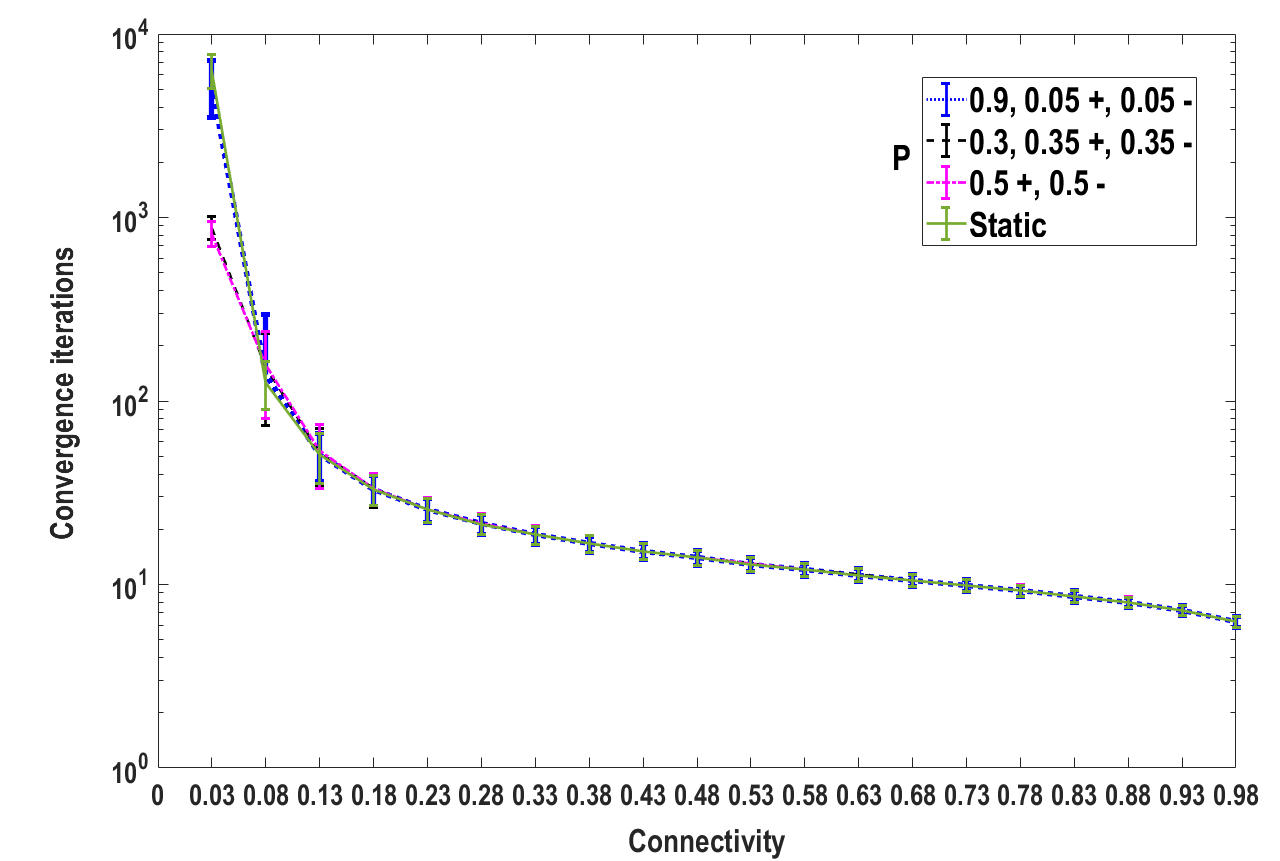}
\caption{Convergence iterations for dynamic networks.}
\label{fig:dynamic}
\end{center}
\end{figure}

We extended Algorithm~\ref{alg:2}'s performance in relation to different values of network connectivity for different values of $P_1$. 
The results presented in Fig.~\ref{fig:dynamic} are achieved in a network consisting of 100 nodes. Results shown on the figure are illustrating the algorithm's performance for $P \in$ \big \langle zero W change, add edge, remove edge \big \rangle. 
In the same case with connectivity r = 0.03, the average convergence iterations is 6400 for the static network, 5312 for $P = \big \langle 0.9, 0.05, 0.05 \big \rangle$, 884 for $P = \big \langle 0.3, 0.35, 0.35 \big \rangle$, and 821 for $P = \big \langle 0.0, 0.5, 0.5 \big \rangle$. 
With $r = 0.08$, the average is 159 for $P = \big \langle 0.0, 0.5, 0.5 \big \rangle$, 154 for  $P = \big \langle 0.3, 0.35, 0.35 \big \rangle$, 140 for $P = \big \langle 0.9, 0.05, 0.05 \big \rangle$, and 126 for static case. For $r \geq 0.13$, the difference between average convergence of dynamic and static networks narrows down.  A key observation of these results is that the convergence depends strongly on the connectivity and the network dynamic. As the probability of dynamic connectivity decreases, and the network connectivity increases, the convergence iterations decreases. 

\subsection{Oscillator drift}

The previous sections established the ability to syntonize the oscillators in the absence of oscillator frequency drift. In practice, the wireless syntonization signals may be implemented periodically, thus the frequencies will be aligned and then drift for some period of time before being aligned again. In this section, the convergence of the decentralized consensus averaging algorithm in the presence of such oscillator drift is analyzed. 

The rate at which the frequency of an oscillator drifts is strongly dependent on the design of the oscillator. A common metric is the Allan deviation (ADEV), which is the square-root of the two-sample variance of the oscillator frequency drift, thus a lower ADEV indicates a more stable oscillator. The ADEV of common temperature-controlled crystal oscillators (TCXOs) is on the order of $1\times10^{-9}$ to $1\times10^{-10}$ over a 1s interval. At short intervals, the frequency drift is heavily influenced by noise fluctuations, and as the interval increases these fluctuations tend to average out, resulting in a decreased ADEV and a more stable frequency. Eventually, long-term drift causes the ADEV to increase. In syntonizing systems in a coherent distributed array, it is necessary to perform coordination updates relatively frequently since the nodes in the array may be in motion. Algorithm~\ref{alg:3} describes the detailed operation of the consensus algorithm with oscillator drift.
\begin{algorithm}
\caption{DFAC with oscillator drift}\label{alg:3}
\begin{algorithmic}
\State {\bf Input:}  $k=0$, the initial estimates $\mathbf{f}(0)$, mixing matrix $\mathbf{W}$
\While{stopping criteria is not satisfied}
    \State $k = k+1$
    \State Generate random vector in $\boldsymbol{\alpha}\in [0,\sigma_\phi]$ from uniform distribution, where $\sigma_\phi$ is the phase error due to random frequency drift
    \State $\mathbf{f}(k-1)= \boldsymbol{\alpha} + \mathbf{f}(k-1)$  
    \State $\mathbf{f}(k)= \mathbf{W}\mathbf{f}(k-1)$ 
\EndWhile

\State \Return $\mathbf{f}(k)$
\end{algorithmic}
\end{algorithm}

For this work, the decentralized consensus algorithm is investigated using a 1s interval time, over which the ADEV is taken to be $1\times10^{-9}$. The frequencies of each node in the array are updated at each processing interval with a frequency drift randomly drawn from a distribution matching the ADEV at a carrier frequency of 1 GHz. The array was set up with a connectivity ratio of 0.15 (or greater if a connectivity of 0.15 was not possible due to the array size being too small). The convergence, standard deviation of the total phase error, and the coherent gain \eqref{Gc} were evaluated for array sizes of 20, 60, and 100 nodes.

\begin{figure}[t!]
\centering
\begin{center}
\end{center}
\begin{subfigure}[b]{0.45\textwidth}
   \includegraphics[width=1\linewidth]{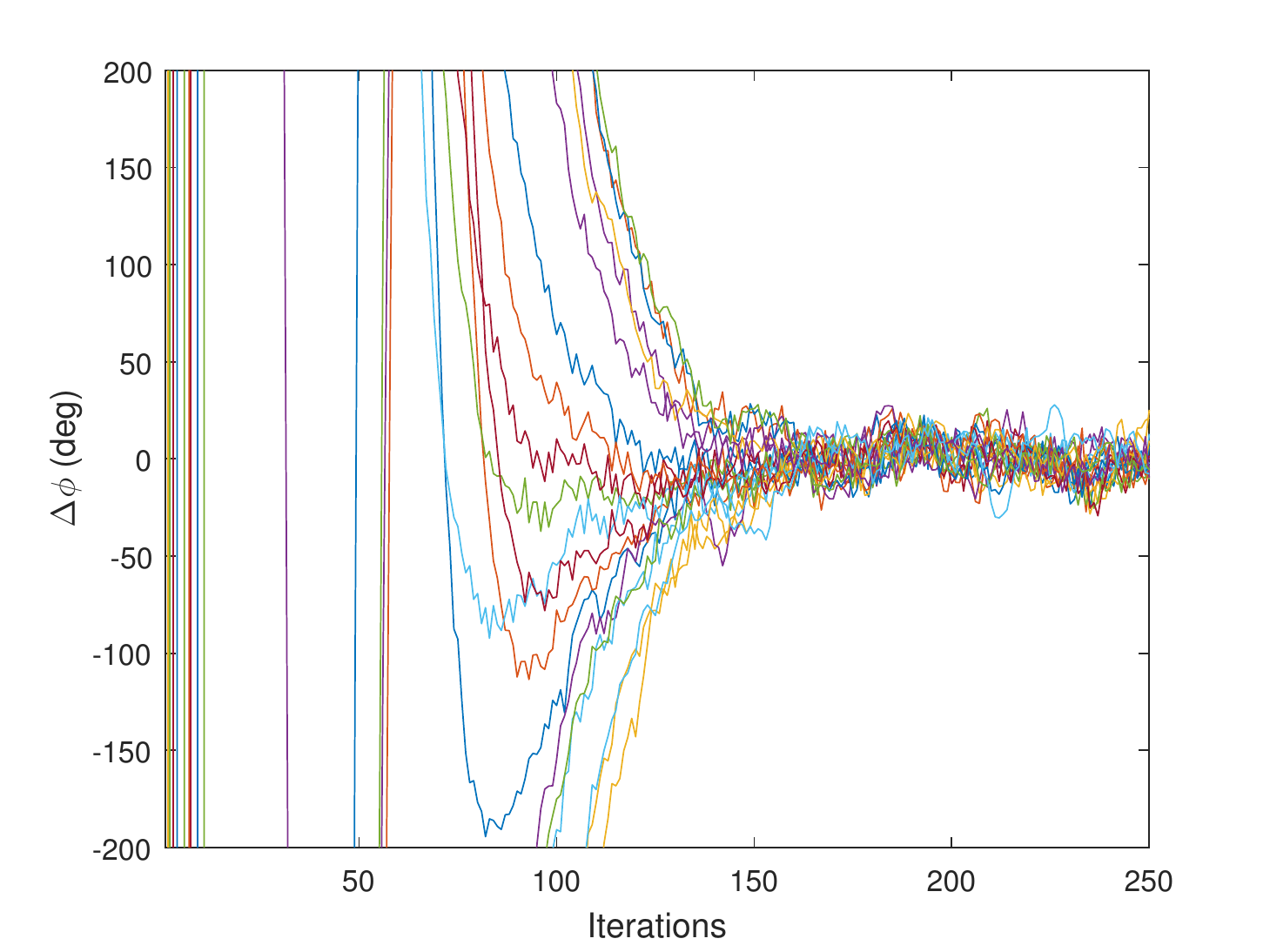}
  \caption{}
   \label{fig:Ng1} 
\end{subfigure}

\begin{subfigure}[b]{0.45\textwidth}
  \includegraphics[width=1\linewidth]{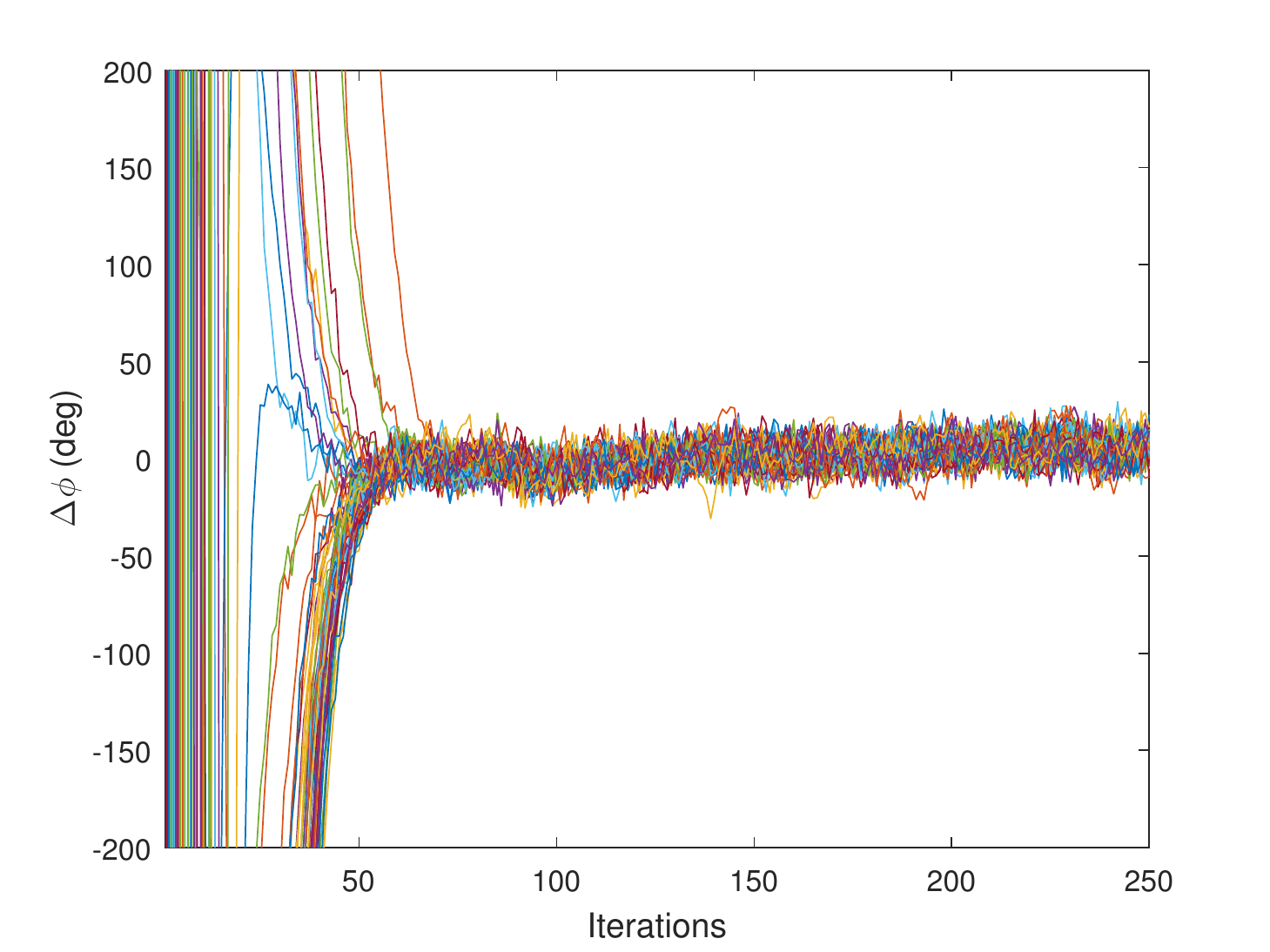}
   \caption{}
   \label{fig:Ng2}
\end{subfigure}

\begin{subfigure}[b]{0.45\textwidth}
  \includegraphics[width=1\linewidth]{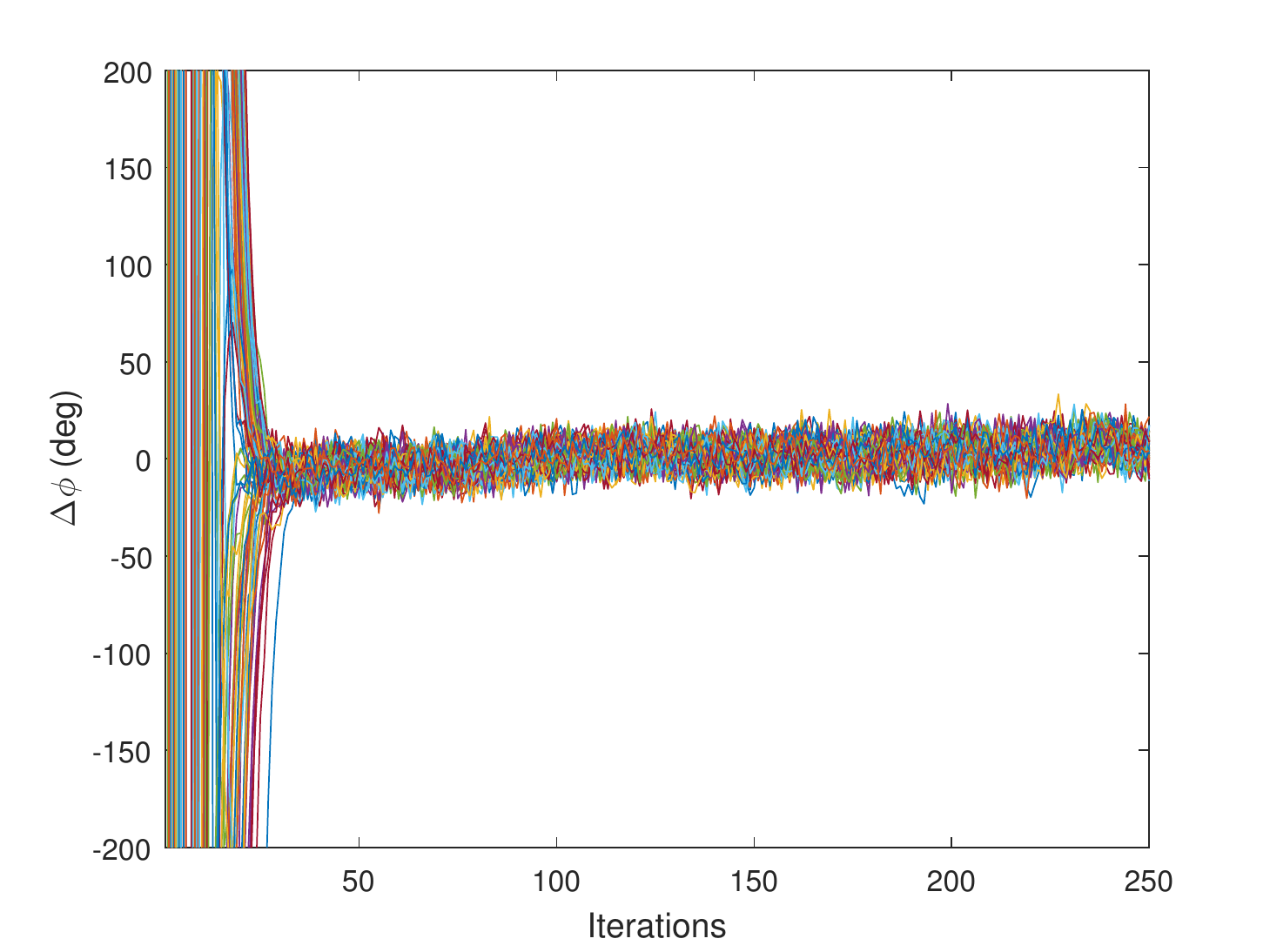}
   \caption{}
   \label{fig:Ng3}
\end{subfigure}
\caption[simulations]{Phases of the oscillators per iteration. (a) $N=20$. (b) $N=60$. (c) $N=100$.}\label{fig:drift}
\end{figure}

\begin{figure}[t!]
\begin{center}
		\noindent
\includegraphics[width=3.6in]{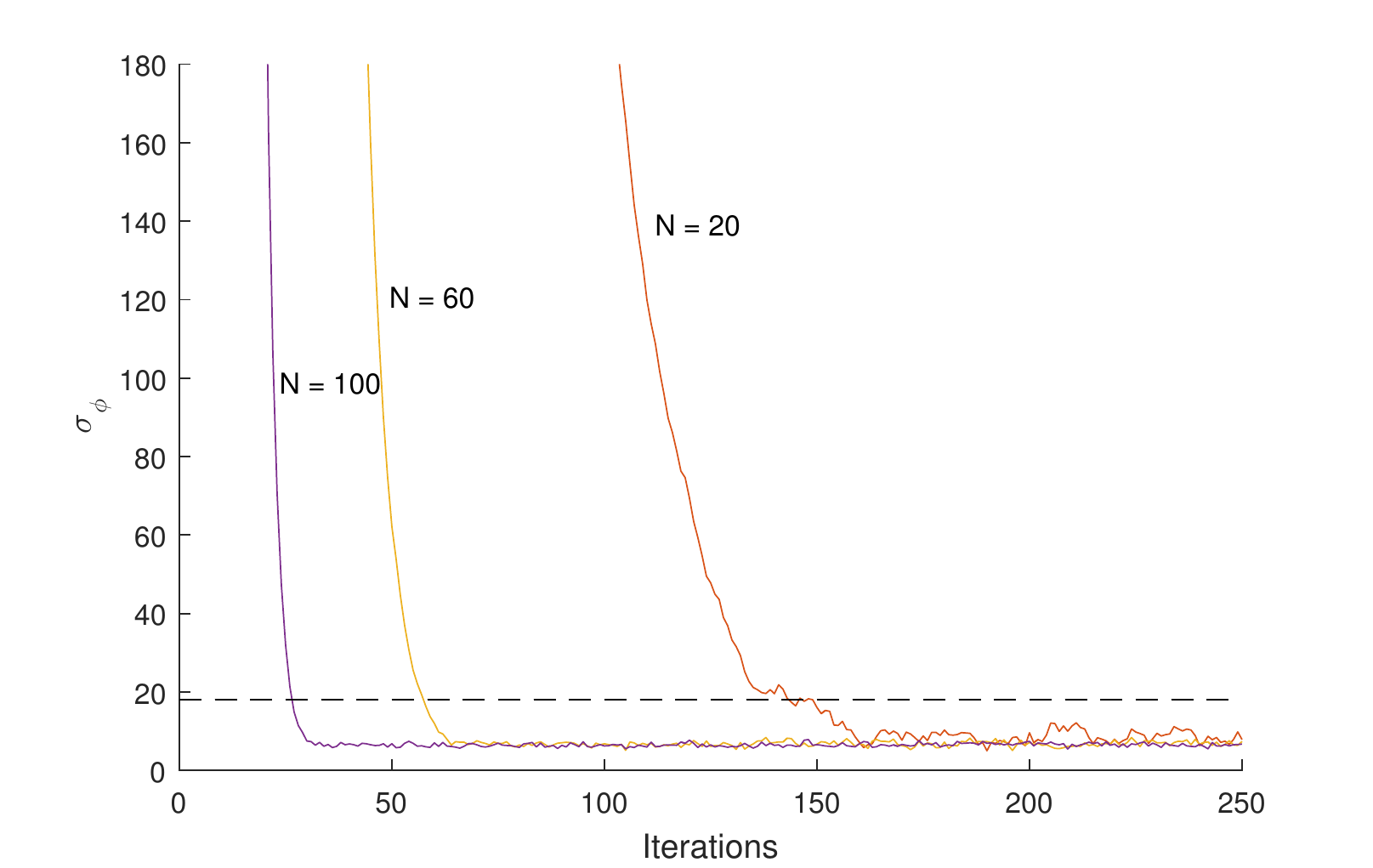}
\caption{Standard deviation of the phases in the arrays. The dotted line indicates the $18\degree$ requirement to achieve 90\% of the ideal coherent gain.}
\label{fig:dev}
\end{center}
\end{figure}

\begin{figure}[t!]
\begin{center}
		\noindent
\includegraphics[width=3.6in]{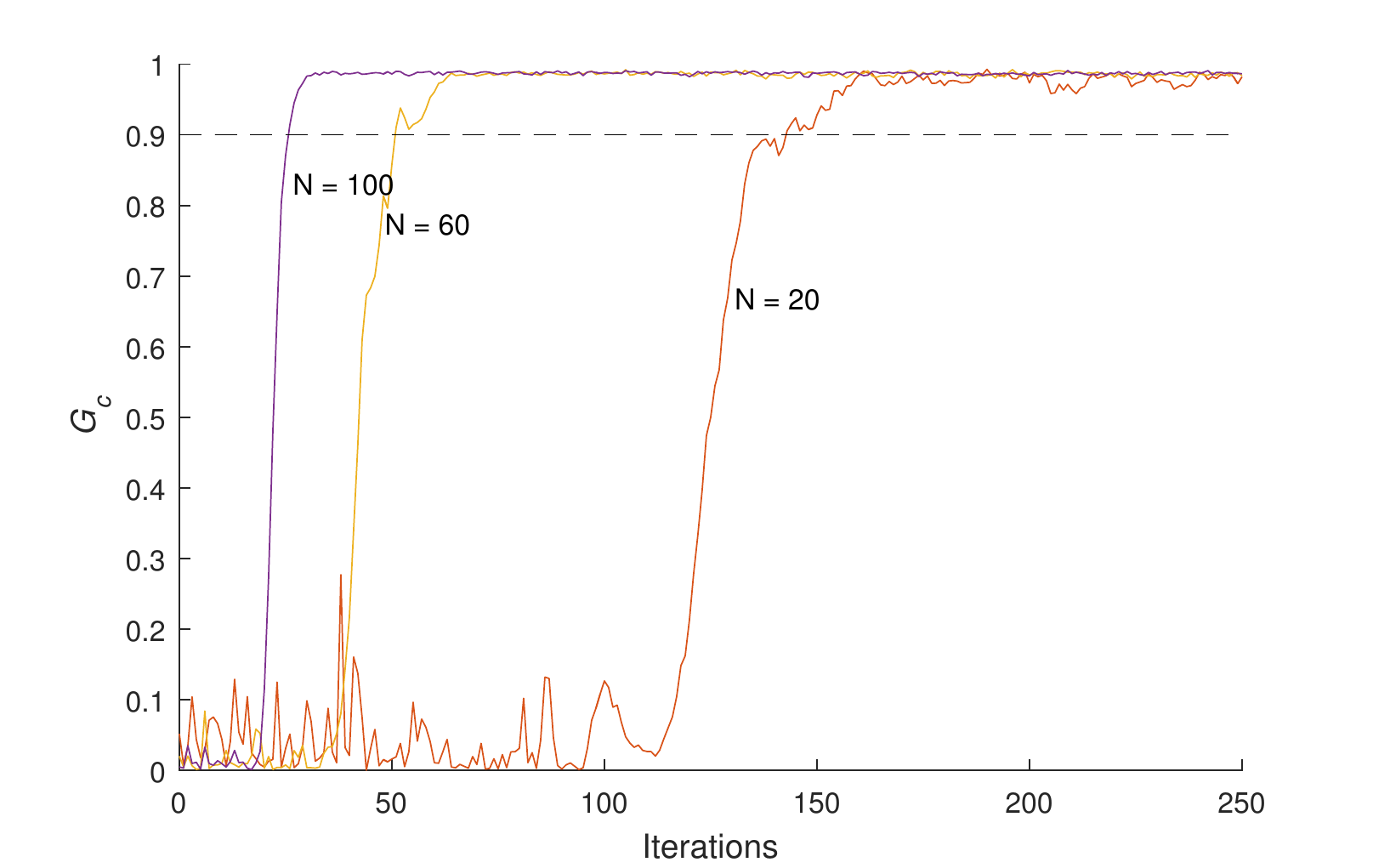}
\caption{Coherent beamforming gain in the antenna mainbeam enabled with the total phase errors given in Fig. \ref{fig:dev}. The dotted line indicates 90\% of the ideal coherent gain.}
\label{fig:gain}
\end{center}
\end{figure}

Fig.~\ref{fig:drift} shows the evolution of the phases for the cases with $N=20$, $N=60$, and $N=100$, respectively. The phase differences are quite large initially for both, but converge to a tight cluster around 130, 55, and 25 iterations, respectively. The standard deviation of these phases determines the level of coherent gain achieved. Fig.~\ref{fig:dev} shows the standard deviation of the phases of all the oscillators in the array as a function of iteration. It can be seen that as the array size increases the convergence time decreases. This is because the connectivity ratio was fixed to be 0.15, meaning that larger arrays enable the information to propagate more quickly through the array. The dotted line indicates the $18\degree$ requirement discussed in Section~\ref{sec:problem}, and clearly the decentralized algorithm enables phase errors below this level even with the relatively poor frequency stability of $1\times10^{-9}$. Fig.~\ref{fig:gain} shows the level of coherent gain supported by the array in the presence of phase errors, showing that in each case the algorithm enables coherent gain levels well above 90\% of the ideal coherent gain, indicated by the dotted line.

\subsection{Discussion}

The numerical results demonstrate that the decentralized consensus protocol can be applied to other topologies that realistically model arrays dynamics, and should allow coherent gain of the distributed beamformer when (a) there is a high probability of link failure and restoration (b) low frequency stability of the oscillators is present. From the above results, it can be concluded that oscillator drift can be considered as the main source of errors in frequency. 
In systems with mobility, minimizing the number of convergence iterations is necessary to reduce energy consumption. This can be achieved by insuring that sufficient communication links are restored.

\section{Conclusion}
Herein, we have proposed a decentralized average consensus method for distributed achieving frequency synchronization, which is vital for distributed beamforming in open loop coherent distributed arrays, without relying on a centralized approach. We have analyzed the performance of the algorithm in undirected (full duplex) networks with static and dynamic topologies. The algorithm is scalable, in that convergence time decreases with the increasing number of participating nodes. The algorithm is found to provide good performance even under adverse conditions such as link failure or oscillator drift. Our ongoing work regards hardware implementation of the promising properties of the suggested protocol and exploration of the decentralized consensus in directed (half-duplex) networks.

\section*{ACKNOWLEDGMENT}

The authors gratefully acknowledge enlightening discussions with Prof. Sandeep Kulkarni, Department of Computer Science and Engineering, Michigan State University. 


%


\ifCLASSOPTIONcaptionsoff
  \newpage
\fi




\bibliographystyle{IEEEtran}
\bibliography{IEEEabrv,group_bib}

\end{document}